%% file: zwcl2341_astroph.tex
\documentclass[useAMS,usenatbib]{mn2e}

\usepackage{graphicx}

\newcommand{\mincir}{\raise -2.truept\hbox{\rlap{\hbox{$\sim$}}\raise5.truept
\hbox{$<$}\ }}
\newcommand{\magcir}{\raise -2.truept\hbox{\rlap{\hbox{$\sim$}}\raise5.truept
\hbox{$>$}\ }}
\newcommand{\siml}{\raise -2.truept\hbox{\rlap{\hbox{$\sim$}}\raise5.truept
\hbox{$<$}\ }}
\newcommand{\simg}{\raise -2.truept\hbox{\rlap{\hbox{$\sim$}}\raise5.truept
\hbox{$>$}\ }}
\newcommand{\be}{\begin{equation}}
\newcommand{\ee}{\end{equation}}
\newcommand{\ba}{\begin{eqnarray}}
\newcommand{\ea}{\end{eqnarray}}
\newcommand {\kpc} {$h_{70}^{-1}$ kpc $\;$}

\newcommand {\hh} {$h_{70}^{-1}$ Mpc}
\newcommand {\hhh} {\;h_{70}^{-1} \mathrm{Mpc}}
\newcommand {\ks} {km~s$^{-1} \;$}
\newcommand {\kss} {km~s$^{-1}$}

\newcommand {\mqui} {$\times 10^{15}\;h_{70}^{-1}\;M_{\odot} \;$}
\newcommand {\mquii} {$\times 10^{15}\;h_{70}^{-1}\;M_{\odot}$}

\newcommand{\degree}{\ensuremath{\mathrm{^\circ}}}

\newcommand{\arcs}{\ensuremath{\arcmm\hskip -0.1em\arcmm \;}}
\newcommand{\arcmm}{\ensuremath{\mathrm{^\prime}}}

\newcommand{\dotsec}{\,\rlap{\hbox{$\mathrm{^s}$}}{\hbox{$.$}}\,}

\title[The dynamical status of ZwCl 2341.1+0000]{The dynamical status
  of ZwCl 2341.1+0000: a very elongated galaxy structure with a
  complex radio emission}

\author[W. Boschin, M. Girardi and R. Barrena]{W. Boschin$^{1,2}$\thanks{E-mail:boschin@tng.iac.es}, M. Girardi$^{2,3}$ and R. Barrena$^{4,5}$\\
$^{1}$Fundaci\'on G. Galilei - INAF (Telescopio Nazionale Galileo),
  Rambla J. A. Fern\'andez P\'erez 7, E-38712 Bre\~na Baja (La Palma),
  Spain\\
$^{2}$Dipartimento di Fisica dell'Universit\`a degli Studi di Trieste
  - Sezione di Astronomia, via Tiepolo 11, I-34143 Trieste, Italy\\
$^{3}$INAF - Osservatorio Astronomico di Trieste, via Tiepolo 11,
  I-34143 Trieste, Italy\\
$^{4}$Instituto de Astrof\'{\i}sica de Canarias, C/V\'{\i}a L\'actea
  s/n, E-38205 La Laguna (Tenerife), Spain\\
$^{5}$Departamento de Astrof\'{\i}sica, Univ. de La Laguna, Av. del
  Astrof\'{\i}sico Francisco S\'anchez s/n, E-38205 La Laguna
  (Tenerife), Spain
}

\begin{document}

\date{Accepted 2013 June 12. Received 2013 June 11; in original form 2013 April 22}

\pagerange{\pageref{firstpage}--\pageref{lastpage}} \pubyear{2013}

\maketitle

\label{firstpage}

\begin{abstract}

We study the dynamical status of the galaxy system ZwCl 2341.1+0000, a
filamentary multi-Mpc galaxy structure associated with a complex diffuse
radio emission.

Our analysis is mainly based on new spectroscopic data for 128
galaxies acquired at the Italian Telescopio Nazionale Galileo. We also
use optical data available in the Sloan Digital Sky Survey and X-ray
data from the {\it Chandra} archive.

We select 101 cluster member galaxies and compute the cluster
redshift $\left<z\right>\sim 0.2693$ and the global line-of-sight
velocity dispersion $\sigma_{V}\sim 1000$ \kss.

Our optical analysis agrees with the presence of at least three,
likely four or more, optical subclusters causing the
south-south-east--north-north-west (SSE--NNW) elongation of the galaxy
distribution and a significant velocity gradient in the south-north
direction. In particular, we detect an important low-velocity subclump
in the southern region, roughly coincident with the brightest peak of
the diffuse radio emission but with a clear offset between the optical
and radio peaks. We also detect one (or two) optical subcluster(s) at
north, in correspondence with the second brightest radio emission, and
another one in the central cluster region, where a third diffuse radio
source has been recently detected. A more refined analysis involving
the study of the 2D galaxy distribution suggests an even more complex
structure. Depending on the adopted model, we obtain a mass estimate
$M_{\rm sys}\sim$ 1-3 \mqui for the whole system.

As for the X-ray analysis, we confirm the SSE--NNW elongation of the
intracluster medium and detect four significant peaks.  The X-ray
emission is strongly asymmetric and offsetted with respect to the
galaxy distribution, thus suggesting a merger caught in the phase of
post--core--core passage.

Our findings support two possible hypotheses for the nature of the
diffuse radio emission of ZwCl 2341.1+0000: a two relics + halo
scenario or diffuse emission associated with the infall and merging of
several galaxy groups during the first phase of the cluster formation.

\end{abstract}


%
%________________________________________________________________

\section{Introduction}
\label{intro}

An increasing fraction of galaxy clusters exhibits diffuse (on Mpc
scale) radio emission which is not associated with the nuclear activity
of member galaxies (e.g., Ferrari et al. \citeyear{fer08}; Venturi
\citeyear{ven11}). The non-thermal nature of these sources and their
steep radio spectra can be interpreted admitting the existence of
relativistic intracluster particles moving in large-scale cluster
magnetic fields.

In the literature these radio sources are usually classified as relics
and haloes. For both types of sources, cluster mergers have been
proposed as the process responsible for their origin (e.g., Feretti
\citeyear{fer99}). Nevertheless, the location and observational
properties of relics and haloes are quite different. Relics are
elongated polarized sources found at the cluster outskirts and are
thought to be connected with shocks occurring during mergers (e.g.,
Ensslin et al. \citeyear{ens98}; Hoeft et al. \citeyear{hoe04}). Radio
haloes, instead, have rounder morphologies, are unpolarized and fill
the central cluster regions. They are probably related to the
turbulent motions of the intracluster medium (ICM) following a merger
(e.g., Cassano et al. \citeyear{cas06}; Brunetti et
al. \citeyear{bru09}), but the detailed scenario is still under
discussion.

Very interestingly, in a few cases diffuse non-thermal radio emission
have been found on even larger scales. An example are the bridges
between relics and haloes observed in a few clusters, e.g. Abell 1656
(Coma; Kim et al. \citeyear{kim89}) and Abell 2744 (Govoni et
al. \citeyear{gov01}). There have also been detections of diffuse
radio sources at large distances from a few clusters, e.g. Abell 2255
(Pizzo et al. \citeyear{piz08}) and Abell 2256 (van Weeren et
al. \citeyear{wer09b}). An intriguing case is also that of the diffuse
source 0809+39 (Brown \& Rudnick \citeyear{bro09}), whose southern
component is possibly associated with a galaxy filament $\sim$5 Mpc
long at $z\sim$0.04. We refer to the recent review by Feretti et
al. (\citeyear{fer12}) for further details and open issues.

\begin{figure*}
\centering 
\includegraphics[width=18cm]{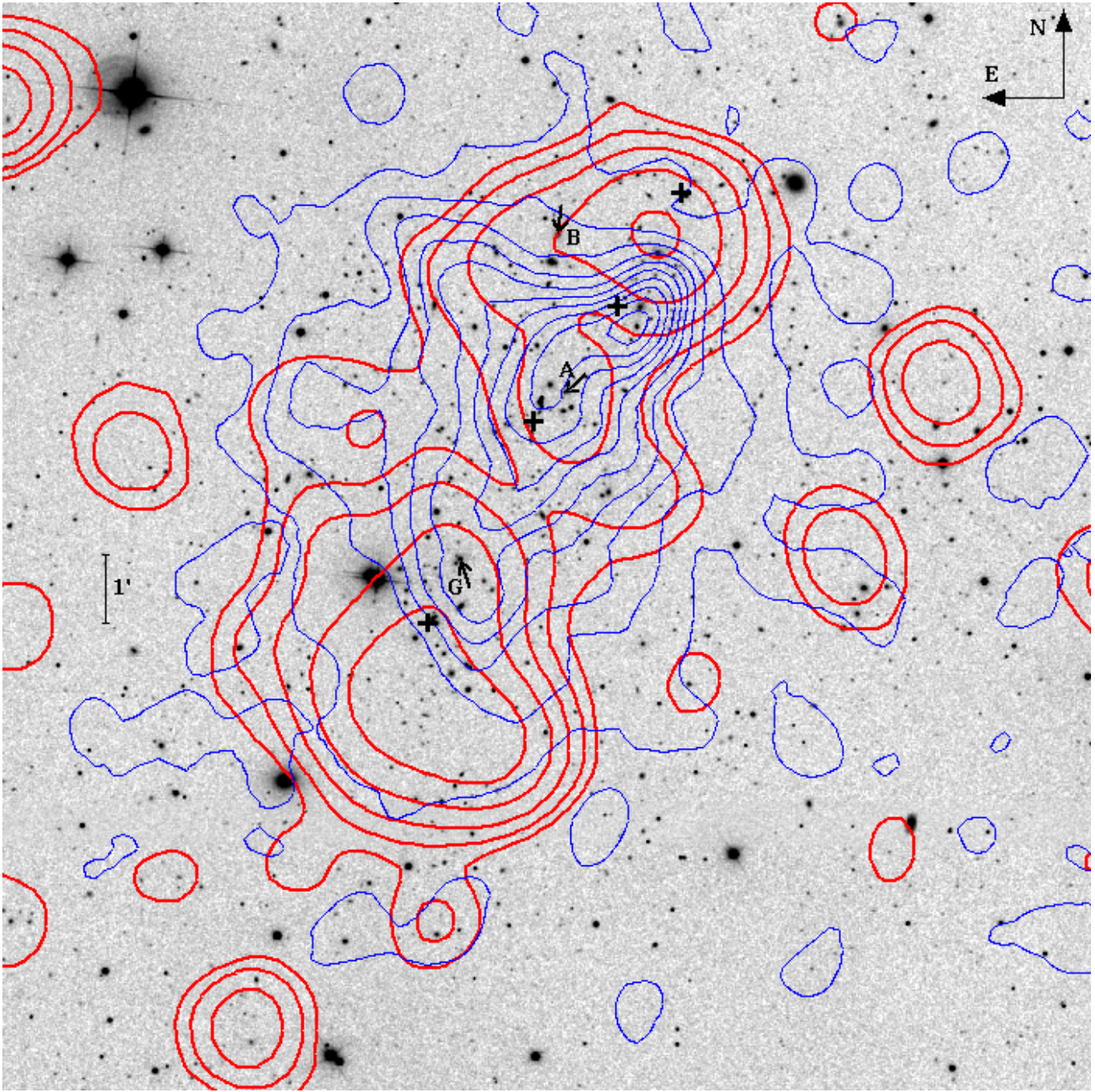}
\caption{SDSS $r^{\prime}$-band image of ZwCl2341+00. Thin (blue)
  contours are the smoothed contour levels of the X-ray surface
  brightness derived from {\it Chandra} archival data (photons in the
  energy range 0.5-2 keV; pointlike sources subtracted; see
  Sect.~\ref{Xmorph}). Thick (red) contours are the contour levels of
  a VLA radio image at 1.4 GHz (discrete sources subtracted, courtesy
  of G. Giovannini). Crosses highlight the centres of the
  substructures detected by the 2D-DEDICA method (see text and
  Fig.~\ref{figk2z}, lower panel). The three arrows and letters
  indicate the head--tail galaxies found by van Weeren et
  al. (\citeyear{wer09a}) and the direction of the radio tails (see
  text).}
\label{figimage}
\end{figure*}

In this paper we focus on a exceptional radio emission associated with
a filamentary multi-Mpc galaxy structure in the ZwCl 2341.1+0000
(hereafter ZwCl2341+00) cluster region, stretching over an area of at
least 4 \hh, as found by Bagchi et al. (\citeyear{bag02}, Very Large
Array, VLA, data).  Bagchi et al. argue that this region is the site
of cosmic shocks originating in multiple mergers during the large
scale structure formation. The radio images obtained from Giant
Metrewave Radio Telescope (GMRT) observations show two diffuse sources
to the north and south of the cluster, interpreted as a double relic,
arising from outgoing shock fronts because of a cluster merger (van
Weeren et al. \citeyear{wer09a}).  In addition to the above relics, ad
hoc VLA observations have allowed Giovannini et al. (\citeyear{gio10})
to detect radio emission in the optical filament of galaxies between
(see Fig.~\ref{figimage}). The size of the radio emission is 2.2 \hh.
In light of their new results, Giovannini et al. discuss three
possible scenarios: the radio emission associated with the cosmic
shocks as suggested by Bagchi et al. (\citeyear{bag02}), a giant radio
halo in-between two symmetric relics, or the merging of two clusters
both hosting a central radio halo. X-ray data from {\it ROSAT}, {\it
  Chandra}, {\it XMM--Newton} and {\it Suzaku} show that ZwCl2341+00
has a complex structure (Bagchi et al. \citeyear{bag02}; van Weeren et
al. \citeyear{wer09a}; Bagchi et al. \citeyear{bag11}; Akamatsu \&
Kawahara \citeyear{aka13}), with a global X-ray temperature $kT_{\rm
  X}\sim 4-5$ keV and luminosity $L_\mathrm{X}$(0.3--8.0 keV)=3$\times
10^{44} \ h^{-2}$ erg\ s$^{-1}$ (van Weeren et
al. \citeyear{wer09a}). The galaxy distribution, as derived from
photometric data of the Sloan Digital Sky Survey (SDSS), is quite
elongated (van Weeren et al. \citeyear{wer09a}). However, no extensive
optical spectroscopy exists for galaxies in the ZwCl2341+00 region and
no dynamical analysis has ever been performed, SDSS data only allowing
to estimate the cluster redshift (z=0.267; Bagchi et
al. \citeyear{bag02}).

\begin{figure*}
\centering 
\includegraphics[width=18cm]{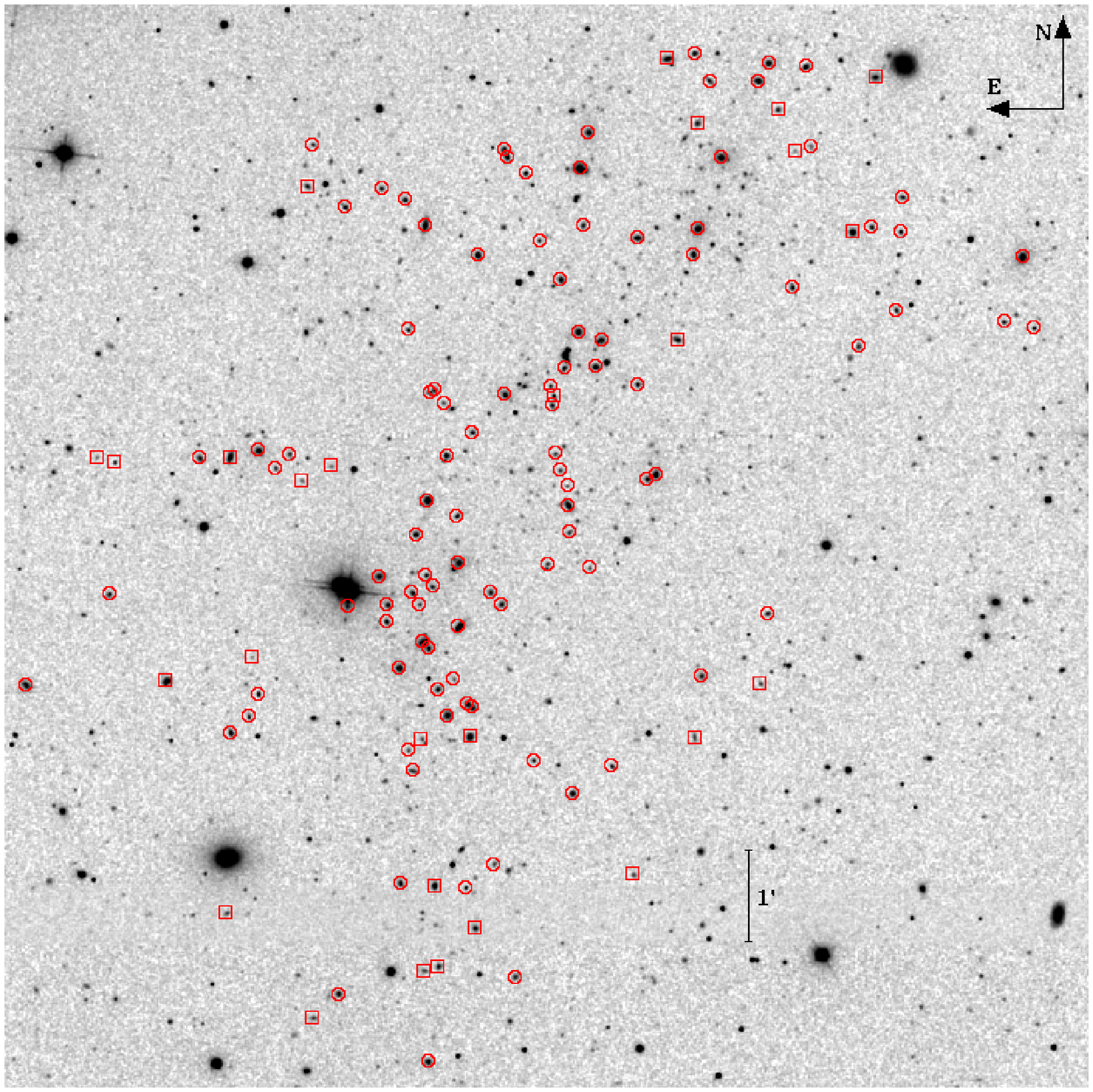}
\caption{SDSS $r^{\prime}$-band image of the cluster
  ZwCl2341+00. Circles and squares indicate cluster members and
  non-member galaxies, respectively (see Table~\ref{catalogZwCl2341}).}
\label{figottico}
\end{figure*}

In the context of our observational program aimed to study the
internal dynamics of clusters with diffuse radio emission (Dynamical
Analysis of Radio Clusters -- DARC; see Girardi et
al. \citeyear{gir10conf}), we performed an intensive study of
ZwCl2341+00. In particular, we obtained new spectroscopic data by
using the Telescopio Nazionale Galileo (TNG). Our present analysis is
mainly based on these optical data. We also used archival optical data
from the SDSS and X-ray data from the {\it Chandra} archive. These
optical and X-ray data complement each other to provide a more
complete picture of the dynamical status of this cluster, as widely
demonstrated by our previous works on other clusters (e.g., Barrena et
al. \citeyear{bar11}; Boschin et al. \citeyear{bos12b}) and also by
the Multiwavelength Sample of Interacting Clusters (MUSIC) project
(Maurogordato et al. \citeyear{mau11} and references therein).

This paper is organized as follows. We present the optical data and
the cluster catalogue in Section~2. We present our results about the
cluster structure based on optical and X-ray data in Sections~3 and 4,
respectively.  We discuss our results and present our conclusions in
Section~5.  Unless otherwise stated, we indicate errors at the 68 per
cent confidence level (hereafter c.l.).  Throughout this paper, we use
$H_0=70$ km s$^{-1}$ Mpc$^{-1}$ and $h_{70}=H_0/(70$ km s$^{-1}$
Mpc$^{-1}$) in a flat cosmology with $\Omega_0=0.3$ and
$\Omega_{\Lambda}=0.7$. In the adopted cosmology, 1 arcmin corresponds
to $\sim 248$ \kpc at the cluster redshift.

\section{Spectroscopic observations}
\label{spec}

Multi-object spectroscopic observations of ZwCl2341+00 were carried
out at the TNG in 2009 October, 2011 August and 2011 December. We used
the instrument Device Optimized for the Low Resolution (DOLORES) in
multi-object spectroscopy (MOS) mode with the LR-B
Grism.\footnote{http://www.tng.iac.es/instruments/lrs} In summary, we
observed four MOS masks for a total of 142 slits. The total exposure
time was 3600 s for three masks and 5400 s for the last one.

Reduction of spectra and radial velocities computation with the
cross-correlation technique (Tonry \& Davis \citeyear{ton79}) were
performed using standard
{\sevensize IRAF}\footnote{{\sevensize IRAF} is distributed by the
  National Optical Astronomy Observatories, which are operated by the
  Association of Universities for Research in Astronomy, Inc., under
  cooperative agreement with the National Science Foundation.}  tasks,
as done with other clusters included in our DARC sample (for a
detailed description see, e.g., Boschin et al. \citeyear{bos12a}). In
seven cases (IDs.~21, 30, 39, 45, 47, 58, and 69; see
Table~\ref{catalogZwCl2341}) the redshift was estimated measuring the
wavelength location of emission lines in the spectra. Our
spectroscopic catalogue lists 128 galaxies in the field of
ZwCl2341+00.

Comparing the velocity measurements for the seven galaxies observed
with multiple masks (see discussion in, e.g., Boschin et
al. \citeyear{bos04}, Girardi et al. \citeyear{gir11}), we corrected
the velocity errors provided by the cross-correlation technique by
multiplying them by a factor of 2.2. Taking into account the above
correction, the median value of the $cz$ errors is 92 \kss.  Three
galaxies have spectroscopic redshifts in the SDSS (Data Release 7), in
good agreement with our values.  With the exception of one galaxy,
SDSS dereddened magnitudes $g^{\prime}$, $r^{\prime}$, and
$i^{\prime}$ are available.

Table~\ref{catalogZwCl2341} lists the velocity catalogue (see also
Fig.~\ref{figottico}): identification number of each galaxy, ID
(column~1); right ascension and declination, $\alpha$ and $\delta$
(J2000, column~2); SDSS (dereddened) $r^{\prime}$ magnitude (column~3);
heliocentric radial velocities, $v=cz_{\sun}$ (column~4) with errors,
$\Delta v$ (column~5).

\subsection{Notable galaxies}
\label{notable}

No galaxy dominates the galaxy population of ZwCl2341+00, with five
galaxies in the brightest magnitude range $17.5<r^{\prime}<18$ (ID~2,
20, 41, 72 and 91).  

The region is populated by a number of radio and X-ray sources. In
particular, van Weeren et al. (\citeyear{wer09a}) took high-resolution
GMRT images at 650 MHz and detected 12 radio sources associated with
individual galaxies. Among them, sources A, B and G are classified as
head--tail galaxies (see Fig.~\ref{figimage}). Their optical
counterparts are member galaxies ID~36, 39 and 73, respectively. Other
unresolved radio sources are C (galaxy ID~41), F (ID~20), H (ID~72), I
(ID~101) and L (ID~95).

Our analysis of {\it Chandra} X-ray data (see Sect.~\ref{Xmorph}) also
detects several X-ray pointlike sources in the field of
ZwCl2341+00. Two of them are galaxies ID~124 (nonmember) and ID~21
(member).

\section{Analysis of the optical data}
\label{anal}

\subsection{Member selection}
\label{memb}

\begin{figure}
\centering
\resizebox{\hsize}{!}{\includegraphics{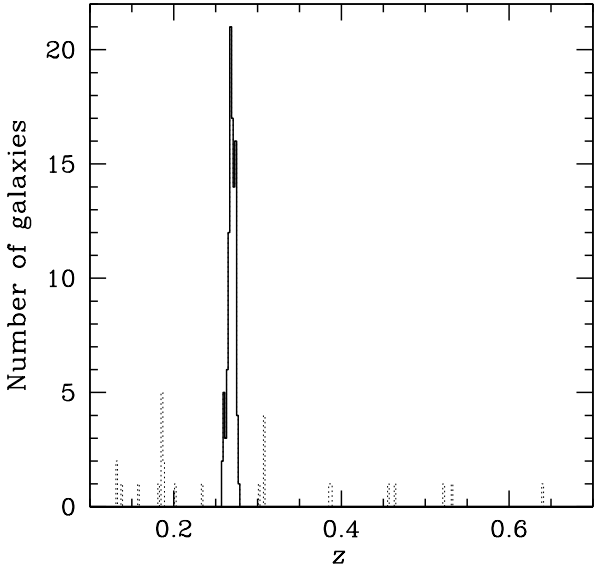}}
\caption
{Redshift galaxy distribution. The solid line histogram refers to the
  101 galaxies assigned to ZwCl2341+00 according to the 1D-DEDICA
  reconstruction method.}
\label{fighisto}
\end{figure}

As usual in the analysis of our DARC clusters, the selection of
cluster members was performed by running two statistical tests. First,
we run the 1D adaptive-kernel method (hereafter 1D-DEDICA; Pisani
\citeyear{pis93} and \citeyear{pis96}) on the 128 galaxies of our
spectroscopic catalogue. This procedure detects ZwCl2341+00 as a peak
at $z\sim0.269$ populated by 101 galaxies considered as fiducial
cluster members (in the range $77\,584\leq v \leq 83\,112$ \kss, see
Fig.~\ref{fighisto}). The 27 non-members are 15 and 12 foreground and
background galaxies, respectively.

In a second step, we combined the spatial and velocity information by
running the `shifting gapper' test by Fadda et al. (\citeyear{fad96};
see also e.g. Girardi et al. \citeyear{gir11} for details). This
procedure confirms the 101 fiducial cluster members selected with the
1D-DEDICA method. We are aware that the results of the `shifting
gapper' can be sensitive to the choice of the cluster centre. Since
ZwCl2341+00 is a complex of three or more groups (see the analysis in
Sect.~\ref{sub}), this choice is not obvious. In this study we adopted
the coordinates of the central galaxy clump as estimated in the
following Sect.~\ref{2d}
[R.A.=$23^{\mathrm{h}}43^{\mathrm{m}}41\dotsec7$, Dec.=$+00\degree
  18\arcmm 13\arcs$ (J2000.0)]. However, we verified that other
possible reasonable choices for the centre, e.g. the highest peak in
the X-ray emission (Bagchi et al. \citeyear{bag02}) or the original
Zwicky centre, do not affect too much the results of the procedure,
leading to the rejection of, at most, two more galaxies.

\begin{figure}
\centering 
\resizebox{\hsize}{!}{\includegraphics{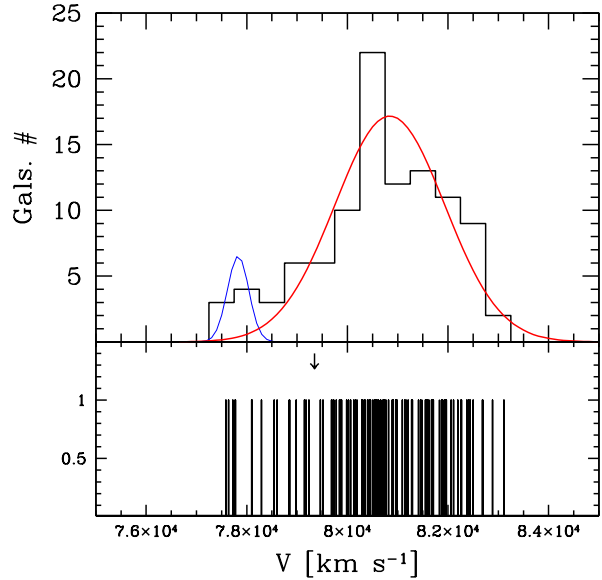}}
\caption
{The 101 galaxies assigned to the cluster. Upper panel: velocity
  histogram. Red (thick) and blue (thin) Gaussians are the best
  bimodal fits according to the 1D-KMM test (see text). Lower panel:
  stripe density plot where the arrow indicates the position of the
  significant gap.}
\label{figstrip}
\end{figure}

\begin{figure}
\centering
\resizebox{\hsize}{!}{\includegraphics{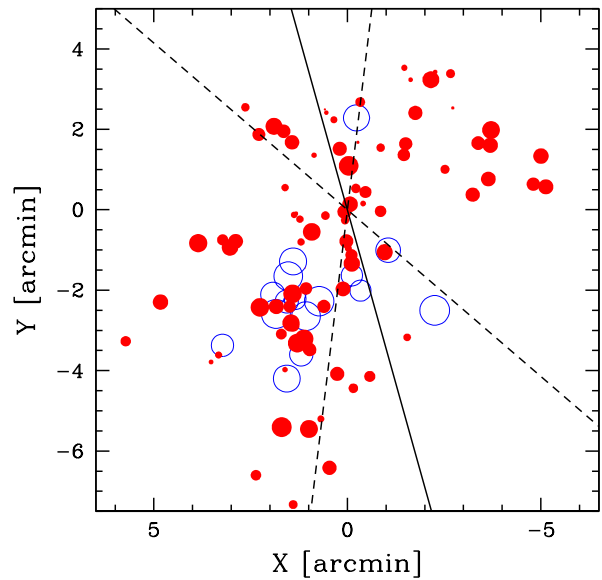}}
\caption
{Spatial distribution on the sky of the 101 cluster members.  The
  larger the symbol, the smaller is the radial velocity. Blue open
  (red full) circles indicate galaxies of the low-velocity group GV1
  (high-velocity group GV2). The plot is centred on the assumed
  cluster centre.  The solid and dashed lines indicate the position
  angle of the cluster velocity gradient (see Sect.~\ref{3d}) and
  relative errors, respectively.}
\label{figgrad}
\end{figure}

\subsection{Global cluster properties}
\label{glob}

The analysis of the velocity distribution of the 101 cluster members
was performed by using the biweight estimators of location and scale
(Beers et al. \citeyear{bee90}). Our measurement of the mean cluster
redshift is $\left<z\right>=0.2693\pm$ 0.0003 (i.e.,
$\left<v\right>=80\,726\pm$103 \kss). As for the second moment of the
velocity distribution (i.e., the global radial velocity dispersion),
our result is $\sigma_{V}=1034_{-74}^{+88}$ \kss.

\subsection{Optical substructures}
\label{sub}

In this subsection we analyse the internal structure of the cluster
and search for eventual substructures. Most of the techniques adopted
in this analysis have already been presented in previous studies of
the DARC clusters and we briefly hint to the reference papers. We only
detail the new developments and the specific results obtained for
ZwCl2341+00.

\subsubsection{Analysis of the velocity distribution of member galaxies}
\label{velo}

The velocity distribution was first analysed to search for possible
deviations from Gaussianity that might provide important signatures of
complex dynamics. In particular, we considered several shape
estimators, e.g. the kurtosis, the skewness and the Scaled Tail Index
(see Bird \& Beers \citeyear{bir93}). We found that the velocity
distribution is significantly skewed (at the 95-99 per cent c.l., see
table~2 of Bird \& Beers \citeyear{bir93}). Moreover, we also applied
the {\it W}-test (Shapiro \& Wilk \citeyear{sha65}) and found a
departure from the Gaussian significant at the 99.6 per cent c.l.

\begin{figure}
%\centering
\includegraphics[width=8cm]{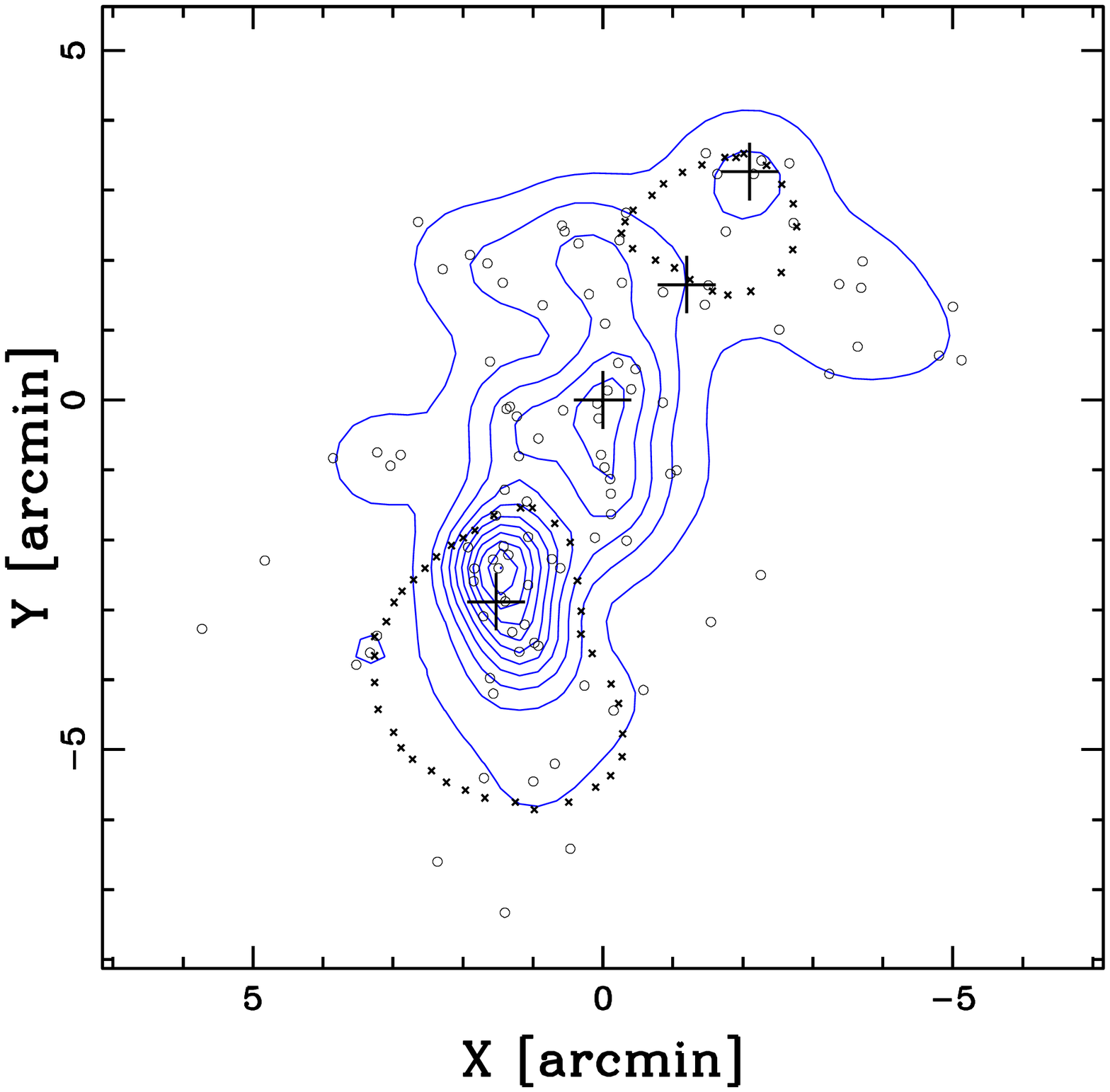}
\includegraphics[width=8cm]{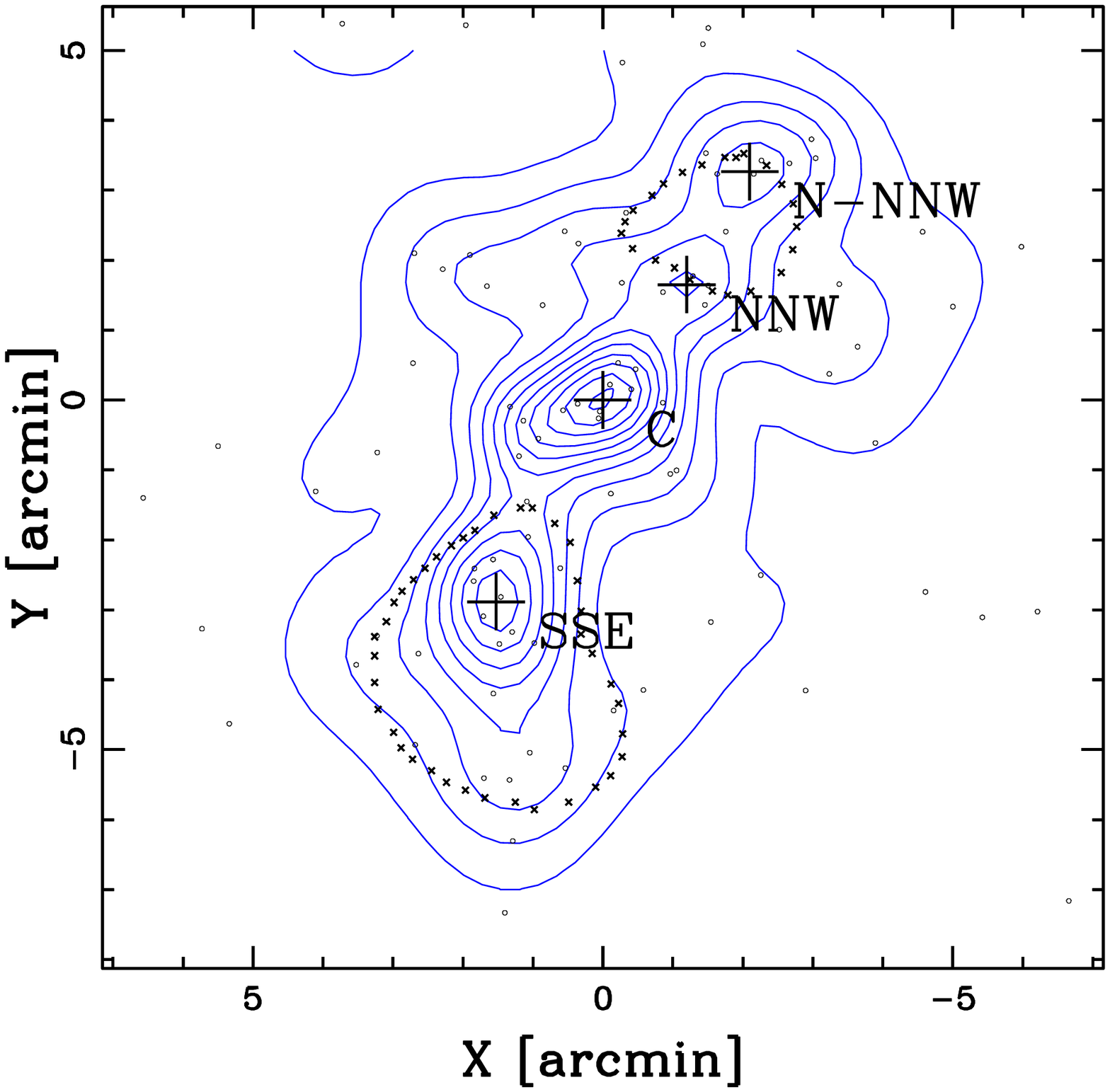}
\caption{Upper panel: spatial distribution on the sky and relative
  isodensity contour map of the 101 spectroscopic cluster members
  obtained with the 2D-DEDICA method. Big crosses indicate the four
  peaks detected in the catalogue of photometric likely members with
  $r^{\prime}\le 20$ (see lower panel). Small crosses indicate, in a
  schematic way, the contour levels of the extended radio emission
  (from fig.~2 -- right hand panel -- of Giovannini et
  al. \citeyear{gio10}, here the second highest contours). Lower
  panel: the same as above for the photometric likely members with
  $r^{\prime}\le 20$ (see text).}
\label{figk2z}
\end{figure}

\begin{figure}
\centering
\resizebox{\hsize}{!}{\includegraphics{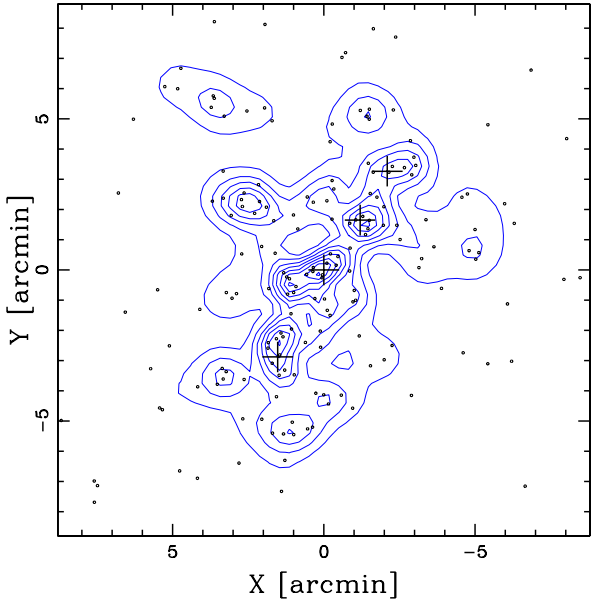}}
\caption{Spatial distribution on the sky and relative isodensity contour map
  of photometric likely cluster members with $r^{\prime}\le 21$. The
  contour map is obtained with the 2D-DEDICA method (blue lines). Big
  crosses indicate the four peaks detected in the catalogue of likely
  members with $r^{\prime}\le 20$ (see Fig.~\ref{figk2z}, lower
  panel).}
\label{figk2m21}
\end{figure}

\begin{figure}
\centering 
\resizebox{\hsize}{!}{\includegraphics{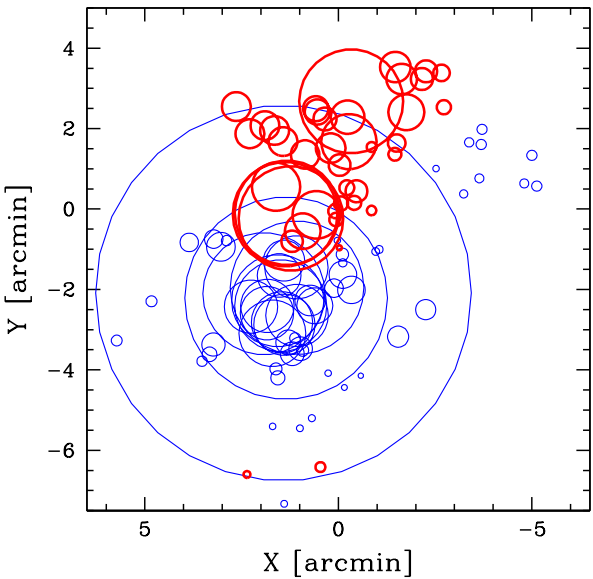}}
\caption
{Spatial distribution of the 101 cluster members, each marked by a
  circle: the larger the circle, the larger is the deviation of the
  local mean velocity from the global mean velocity.  Thin/blue and
  thick/red circles show where the local value is smaller or larger
  than the global value.}
\label{figds10v}
\end{figure}

We then investigated the presence of gaps in the velocity distribution
performing the weighted gap analysis (Beers et al. \citeyear{bee91};
\citeyear{bee92}). We detect one significant gap (at the 97 per cent
c.l.), which divides the cluster into two groups, GV1 and GV2, of 16
and 85 galaxies from low to high velocities, respectively (see
Fig.~\ref{figstrip}).

The 1D-Kaye's mixture model test (1D-KMM; Ashman et
al. \citeyear{ash94}), which fits a user-specified number of Gaussian
distributions to a data set and assesses the improvement of that fit
over a single Gaussian (see, e.g., Boschin et al. \citeyear{bos12a}
for details), also detects a significant two-group partition (at the
98.7 per cent c.l.; seven and 94 galaxies). For the two groups
(hereafter KMM1D-LV and KMM1D-HV) we obtain mean velocities,
$\left<v_{\rm KMM1D-LV}\right>=77\,827$ \ks and $\left<v_{\rm
  KMM1D-HV}\right>=80\,843$ \kss, and velocity dispersions,
$\sigma_{V,\rm KMM1D-LV}=152$ \ks and $\sigma_{V,\rm KMM1D-HV}=857$
\kss, where the galaxies are weighted according to their partial
membership to both the groups. The weights help to avoid possible
underestimates due to an artificial truncation of the tails of the
velocity distributions.  A three-group partition of 15, 53 and 33
galaxies also provides a fit better than the single Gaussian, but only
at a 95 per cent c.l., thus it was not considered according to Bird
(\citeyear{bir94}).

Both GV1 and KMM1D-LV are two low-velocity groups concentrated in the
SE cluster region (see Fig.~\ref{figgrad}). According to the 2D
Kolmogorov--Smirnov test (Fasano et al. \citeyear{fas87}) the spatial
distributions of the galaxies of GV1 and GV2 (KMM1D-LV and KMM1D-HV)
are different at the 99.4 per cent c.l. (97 per cent c.l.). This
result suggests that the cluster structure can be revealed by the
analysis in the phase space (i.e., combining velocities and sky
positions; see Sect.~\ref{3d}).

\subsubsection{Analysis of the spatial distribution of member galaxies}
\label{2d}

We analysed the spatial distribution of the 101 member galaxies by
using the 2D adaptive-kernel method (hereafter 2D-DEDICA). We
highlight a south-south-east--north-north-west (SSE-NNW) elongated
structure with three very significant galaxy peaks (Fig.~\ref{figk2z},
upper panel), the SSE one being the most important.

\begin{figure}
\centering
\resizebox{\hsize}{!}{\includegraphics{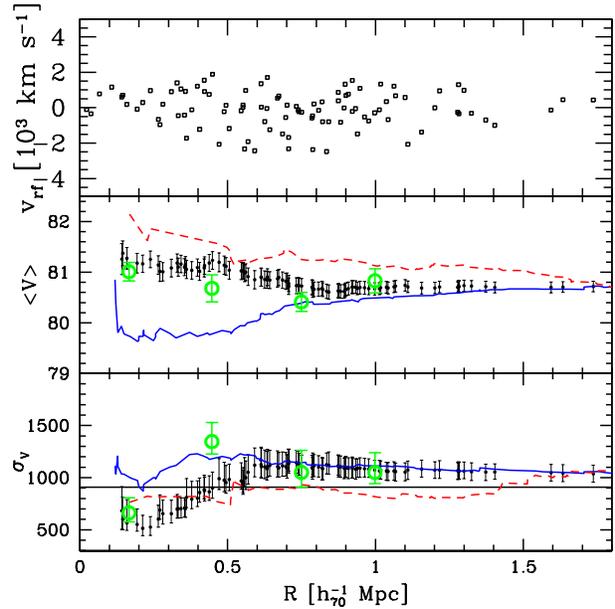}}
\caption
{Top panel: rest-frame velocity versus projected clustercentric
  distance for the 101 spectroscopic member galaxies. Here the centre
  is the `C' peak, assumed as cluster centre in this study. Middle and bottom panels: differential (big green circles) and
  integral (small points) profiles of mean velocity and LOS velocity
  dispersion, respectively.  For the differential profiles, we plot
  the values for four annuli from the centre of the cluster, each of
  0.3 \hh .  For the integral profiles, the mean and dispersion at a
  given (projected) radius from the cluster centre is estimated by
  considering all galaxies within that radius -- the first value
  computed on the five galaxies closest to the centre. The error bands
  at the 68 per cent c.l. are also shown. Blue solid lines (red dashed
  lines) refer to integral profiles assuming the `SSE' peak
  (`N-NNW' peak) as centre.  In the bottom panel, the horizontal
  line represents the value of the X-ray temperature (5 keV) assuming
  the density--energy equipartition between ICM and galaxies.}
\label{figprof}
\end{figure}

\begin{figure}
\centering
\resizebox{\hsize}{!}{\includegraphics{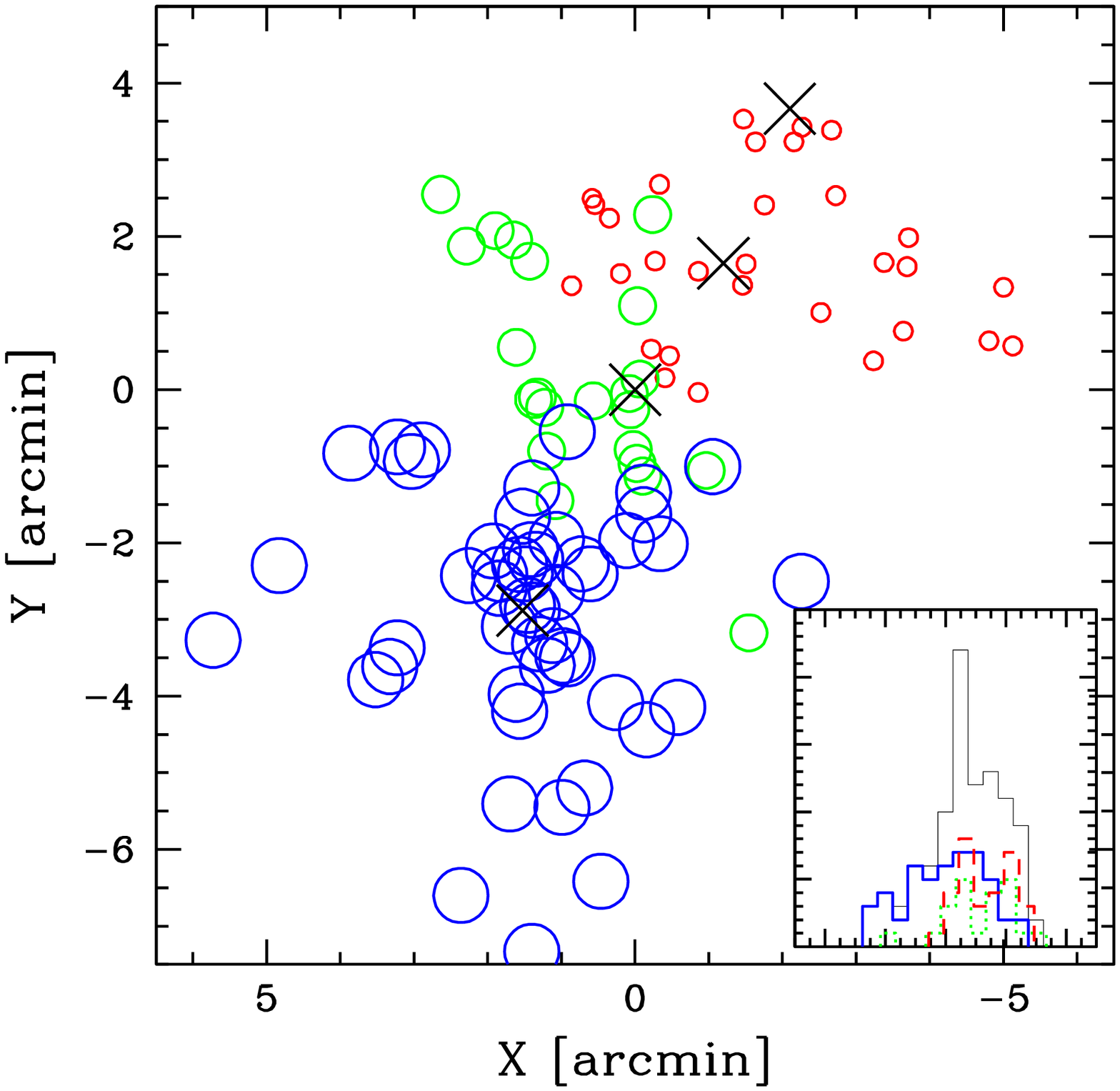}}
\caption{Spatial distribution on the sky of the 101 cluster members.
  Large blue, medium green, and small red circles indicate the
  spectroscopic members belonging to the 3D-KMM groups with increasing
  mean velocities (KMM3D-S-LV, KMM3D-C-HV and KMM3D-N-HV).  The
  crosses are the main galaxy peaks detected in the 2D galaxy
  distribution (see Fig.~\ref{figk2z}, lower panel).  The insert plot
  shows the velocity distributions of the 3D-KMM groups (solid blue
  line, dotted green line, and dashed red line). The insert plot has
  the same axis and units of Fig.~\ref{figstrip} and also shows the
  velocity distribution of all the 101 spectroscopic members for an
  easy comparison.}
\label{figkmm3d}
\end{figure}

Our spectroscopic data do not cover the entire cluster field and are
affected by magnitude incompleteness due to unavoidable constraints in
the design of the MOS masks. In particular, comparing the photometric
and spectroscopic samples in the sky region explored by the four MOS
masks we estimate that we retrieved spectroscopic redshifts for the
$\sim$75 per cent ($\sim$60 per cent) of the galaxies with
$r^{\prime}\le$19.0 ($\le$20.0). To overcome the incompleteness of the
spectroscopic sample we resorted to the photometric SDSS catalogues
extracted in a 30 arcmin radius region from the cluster centre. In
particular, we selected likely members on the basis of both
($r^{\prime}$--$i^{\prime}$ versus $r^{\prime}$) and
($g^{\prime}$--$r^{\prime}$ versus $r^{\prime}$) colour-magnitude
relations (CMRs), which indicate the early-type galaxies locus
(e.g. Dressler \citeyear{dre80}; see Boschin et al. \citeyear{bos12a}
for details on the technique used for the determination of the CMRs
and the selection of member galaxies). The equations of the two CMRs
are $r^{\prime}$--$i^{\prime}$=1.033-0.027$\,r^{\prime}$ and
$g^{\prime}$--$r^{\prime}$=2.523-0.056$\,r^{\prime}$.

Fig.~\ref{figk2z} (lower panel) shows a zoom of the contour map for
the likely cluster members with $r^{\prime}\le 20$ in the region
comparable to that sampled by the spectroscopic data: the results of
the spectroscopic sample are confirmed, with the difference that the
main peak is now the central one (C peak) and the presence of the
additional peak NNW (other peaks shown in the figure are SSE and N-NNW
ones). Fig.~\ref{figk2m21} shows the results for the catalogue of
likely members with $r^{\prime}\le 21$: it shows an even more complex
structure with several subclumps. Two external peaks, corresponding to
other galaxy clusters are outside of the boundaries of the figure.
Table~\ref{tabdedica2d} lists information for the peaks obtained from
the photometric samples: the number of assigned members, $N_{\rm S}$
(column~2); the peak position (column~3); the density (relative to the
densest peak), $\rho_{\rm S}$ (column~4); the value of $\chi^2$ for
each peak, $\chi ^2_{\rm S}$ (column~5).  We list all the very
significant peaks (at $>99.9$ per cent c.l.) with relative density
$\rho_{\rm S} \ge 0.5$.

\subsubsection{3D analysis: combining velocity and spatial information}
\label{3d}

Here we present our analysis of the structure of ZwCl2341+00 combining
the spatial and velocity information of the catalogue of spectroscopic
cluster members. Several approaches were explored.

\begin{figure}
\centering 
\resizebox{\hsize}{!}{\includegraphics{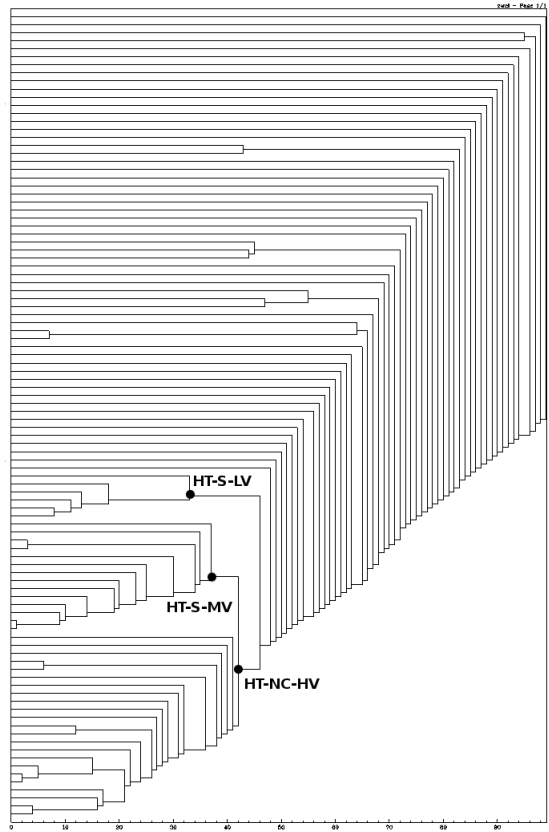}}
\caption{Dendogram obtained through the Serna \& Gerbal
  (\citeyear{ser96}) algorithm. The abscissa is the binding energy
  (here in arbitrary units, with the deepest negative energy levels on
  the left). The three subclusters HT-S-LV, HT-S-MV and HT-NC-HV (see
  text) are indicated by their respective labels.}
\label{zwclgerbal}
\end{figure}

\begin{figure}
\centering
\resizebox{\hsize}{!}{\includegraphics{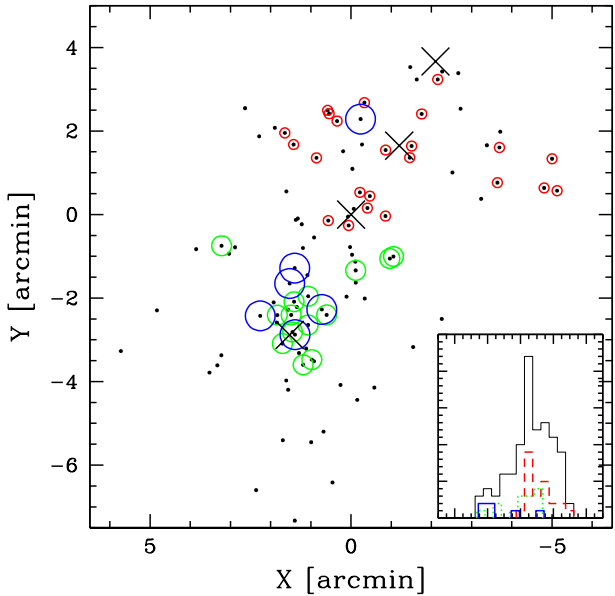}}
\caption{Spatial distribution on the sky of the 101 cluster members
  (black points).  Large blue, medium green, and small red circles
  indicate galaxies assigned to Htree groups with increasing mean
  velocities (HT-S-LV, HT-S-MV and HT-NC-HV). The insert plot shows
  their velocity distributions (solid blue line, dotted green line
  and dashed red line, respectively).}
\label{fight}
\end{figure}

First, we checked for a velocity gradient in the set of the 101
cluster members (see e.g. den Hartog \& Katgert \citeyear{den96} and
Girardi et al. \citeyear{gir96}). We found a significant velocity
gradient (at the 98 per cent c.l.) with a position angle of
PA$=16_{-32^{\rm o}}^{+34^{\rm o}}$ (measured counter clock wise from
north, see Fig.~\ref{figgrad}). This means that higher velocity
galaxies lie at the north and lower velocity galaxies at the south
(see also Fig.~\ref{figds10v}).

Then, we applied the classical $\Delta$-statistics (Dressler \&
Schectman \citeyear{dre88}, hereafter DS-test) to detect the presence
of substructure. In particular, we considered the modified version of
this test which splits the contribution of the local mean velocity
(DS$\left<v\right>$-test) and dispersion (DS$\sigma_{V}$-test; see
for details, e.g., Girardi et al. \citeyear{gir97}; Ferrari et
al. \citeyear{fer03}; Girardi et al. \citeyear{gir10}). The DS-test
detects substructure significant at the 99.9 per cent c.l., mainly due
to the mean velocity estimator (for the DS$\left<v\right>$-test the
significance is at the 99.9 per cent c.l.).

We also analysed the profiles of mean velocity and velocity dispersion
using as centres the C, SSW and N-NNW peaks, which are detected by the
2D analysis of both the spectroscopic and photometric samples
(Sect.~\ref{2d}).  The integral profiles are flat within the errors in
the external regions (see Fig.~\ref{figprof}), thus suggesting that we
obtain a robust value of $\sigma_{V}$ for the whole cluster. However,
the values of the mean velocity and the velocity dispersion depend on
the galaxy peak assumed as the cluster centre.  The `C' subcluster is
well detected as a group with $\sigma_{V}\sim 500$-600 \ks and
$\left<v\right>\sim 81\ 000$ \ks within 0.3 \hh, then $\sigma_{V}$
quickly increases due to the inclusion of other subclusters. As for
the `SSE' and `N-NNW' subclusters, they show lower and higher mean
velocities, respectively.  We also used the galaxies around the three
galaxy peaks (i.e. within 0.25 \hh) as seeds to check for a
significant three-group partition through the 3D-KMM test. We find
that a partition of 33, 22 and 49 galaxies (hereafter KMM3D-N-HV,
KMM3D-C-HV, KMM3D-S-LV groups) is a significantly more accurate
description of the 3D galaxy distribution than a single 3D Gaussian
(at the $>99.9$ per cent c.l.).  The groups are well separated in the
plane of the sky and the southern group (KMM3D-S-LV) is characterized
by a lower mean velocity (see Fig.~\ref{figkmm3d}).

We also resorted to the Htree method devised by Serna \& Gerbal
(\citeyear{ser96}; see also Girardi et al. \citeyear{gir11} for a
recent application and Serra \& Diaferio \citeyear{ser13}, where a
similar technique is also used to determine cluster members). The
method uses a hierarchical clustering analysis to determine the
relationship between galaxies according to their relative binding
energies. Fig.~\ref{zwclgerbal} shows the resulting dendogram, where
three subclusters are identified: one occupying the north and central
regions, characterized by a high mean velocity (HT-NC-HV), and two at
the south with lower mean velocities (HT-S-LV and HT-S-MV; see
Fig.~\ref{fight}).

\begin{figure}
\centering
\resizebox{\hsize}{!}{\includegraphics{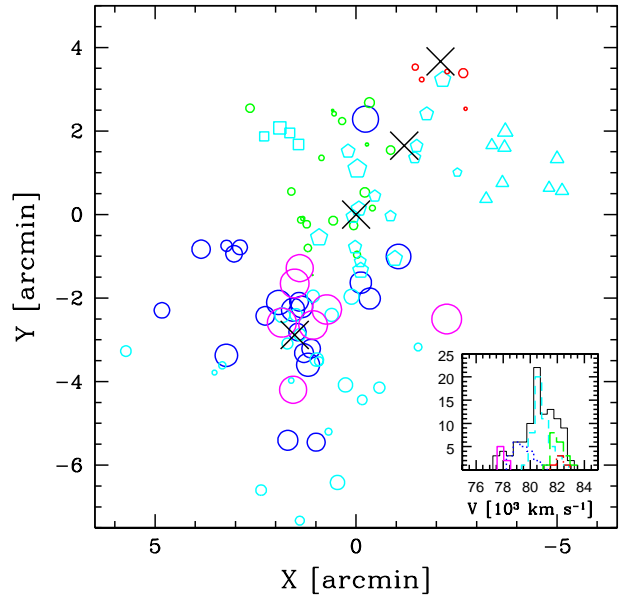}}
\caption{Spatial distribution on the sky of the 101 cluster members.
  Colours magenta, blue, cyan (triangles, squares, pentagons, circles),
  green and red identify galaxies assigned to 3D-DEDICA groups with
  increasing mean velocities (see Table~\ref{tabsub}). The smaller the
  symbol the higher the velocity of individual galaxies. The insert
  plot shows the velocity distributions of the 3D-DEDICA groups (solid
  magenta line, dotted blue line, dashed cyan line, long-dashed green
  line and dot-dashed red line).}
\label{figded3d}
\end{figure}

Finally, we applied the 3D adaptive-kernel method (hereafter
3D-DEDICA; Pisani \citeyear{pis96}, see Girardi et
al. \citeyear{gir11} for a recent application). This technique detects
eight groups (see Fig.~\ref{figded3d}). Among them, the two groups
with the lowest mean velocity lie at south (DED3D-S-LLV, DED3D-S-LV),
and the two ones with the highest mean velocity lie at north and
in the central region (DED3D-N-HHV, DED3D-NC-HV). There are also four
groups with similar mean velocities, all closely around the cluster
mean velocity: we join them together in a single subcluster
(DED3D-NCS-MV) in the insert figure of Fig.~\ref{figded3d} and in
Tab.~\ref{tabsub}. Table~\ref{tabsub} lists the properties of the
subclusters detected by the 3D-DEDICA technique and the methods
described above.

\subsection{Computation of the cluster mass}
\label{mass}

As for the determination of global virial quantities, we followed the
prescriptions of Girardi \& Mezzetti (\citeyear{gir01}, see also
Girardi et al. \citeyear{gir98}) considering different models for
ZwCl2341+00, from a single spherical system to the sum of several
subsystems using the values of the velocity dispersions listed in
Tab.~\ref{tabsub}.  For the single system, we obtain $M_1(<R_{\rm
  vir,1}=2.2 \hhh)=1.5_{-0.4}^{+0.5}$ \mquii. For other models, each
based on a different substructure test, we sum the mass values
obtained for each subcluster but with a rough rescaling with the
respective $R_{\rm vir}$ (i.e., $R_{\rm vir,1}/R_{\rm vir,subcl}$) in
such a way to estimate the mass for a corresponding similar 2.2
\hh-radius region. We find: $M_{\rm 1D-KMM}\sim1$ \mquii, $M_{\rm
  3D-KMM}\sim3$ \mquii, {\bf $M_{\rm Htree}\sim3$ \mquii}, $M_{\rm
  3D-DEDICA}\sim1$ \mquii. The cluster is thus confirmed to be quite
massive, although the precise mass value is uncertain, depending on
the adopted model.

\section{Analysis of the X-ray data}
\label{Xmorph}

In this work we are mainly interested in comparing the spatial galaxy
distribution with the morphology of the ICM. To this aim, we
considered archival X-ray data taken with {\it Chandra} Advanced CCD
Imaging Spectrometer (ACIS-I; exposure ID \#5786, total exposure time
30.2 ks). We reduced the data using the package {\sevensize
  CIAO}\footnote{see http://asc.harvard.edu/ciao/} (version 4.2) on
chips I0, I1, I2, and I3 (field of view $\sim 17\times 17$
arcmin$^{2}$) in a standard way (see e.g. Boschin et
al. \citeyear{bos04}).

\begin{figure}
\centering
\resizebox{\hsize}{!}{\includegraphics{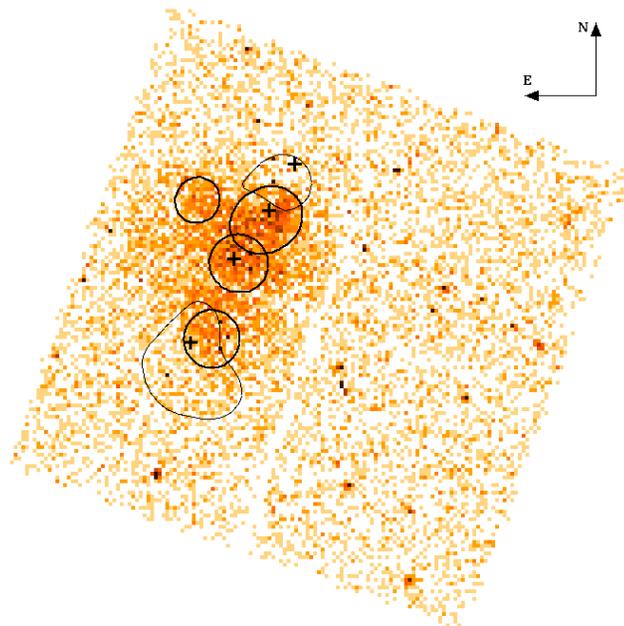}}
\caption{{\it Chandra} X-ray image of the cluster ZwCl2341+00 in the
  energy band 0.5--2 keV (field of view 17$\times$17
  arcmin$^2$). Ellipses are the substructures found by the wavelet
  analysis. Thin contours are the contour levels of the diffuse radio
  emission (here the second highest contours). As a reference, crosses
  indicating the optical peaks detected by 2D-DEDICA (sample of likely
  members with $r^{\prime}\le20$) are also shown.}
\label{wavelet}
\end{figure}

The photon counts of the reduced image are affected by the poorly
exposed ACIS-I CCD gaps (see Fig.~\ref{wavelet}). We corrected this
effect by binning the reduced image, applying a soft smoothing and,
finally, dividing by an exposure map. The result is an image whose
contour levels are plotted in Fig.~\ref{figimage}. A simple look at
the X-ray contours shows the complex morphology of the ICM, which
appears elongated in the SSE--NNW direction, i.e. the direction
defined by the spatial distribution of member galaxies (see
Sect.~\ref{2d}).

A quantitative analysis of the ICM morphology can be performed by
executing a multiscale wavelet analysis. We consider the counts in the
soft band (0.5--2 keV) corrected for the exposure map. After removal
of point-like sources, we run the task {\sevensize CIAO/wavdetect} and
detect four substructures (see Fig.~\ref{wavelet} and
Table~\ref{tabwavelet}). Three of the four X-ray peaks follow the
SSE--NNW direction defined by the optical peaks, with Wav-NNW and Wav-C
being the most significant ones. The structure Wav-NE seems an ICM
filament pointing to the location of the optical NE peak detected by
2D-DEDICA (see Fig.~\ref{figk2m21} and Table~\ref{tabdedica2d}).

As for the spectral properties of the ICM, van Weeren et
al. (\citeyear{wer09a}) and Bagchi et al. (\citeyear{bag11}) have
already measured the overall temperature by analysing both the {\it
  Chandra} data considered in this work and {\it XMM--Newton}
data. Moreover, Akamatsu \& Kawahara (\citeyear{aka13}) present a
temperature profile based on deeper {\it Suzaku} data. We refer to
their results in the discussion. Because of the relatively low photon
count statistics (in particular in the peripheral regions of the ICM)
of {\it Chandra} and {\it XMM-Newton} data we omit the computation of
an accurate temperature map, although it would be desirable to
characterize the X-ray properties of the detected substructures.

\section{Discussion and conclusions}
\label{disc}

We confirm that ZwCl2341+00 has a complex structure. With respect to
the previous detection of a northern and a central peaks in the galaxy
distribution (see fig.~1 of van Weeren et al. \citeyear{wer09a}), we
also detect an important peak at south.  Using our spectroscopic
sample we can detect these galaxy clumps in the 3D phase space, in
particular it is quite clear the presence of a southern low-velocity
structure and a northern high-velocity one. Depending on the adopted
method for the analysis, the cluster structure is less or more
complex, going from a simple two-clump structure in the velocity space
to several clumps in the 3D-DEDICA analysis. Several results support
the existence of at least three structures, at north and south and in
the central regions. The projected cluster structure is strongly
elongated along the SSE--NNW direction, and the angle between the
cluster axis and the plane of sky is probably intermediate, as
suggested by the fact that we detect the subclusters both in the
velocity and in the 2D space.

The ICM X-ray emission roughly follows the distribution of the cluster
galaxies. However, it is strongly asymmetric, with the X-ray peak
located at north with respect to the central (and densest) galaxy
group. Moreover, there is no one-to-one correspondence between the
optical subclumps and the X-ray structures detected by the wavelet
analysis (Table~\ref{tabwavelet}). We deduce that ZwCl2341+00 is
caught during or after the merging phase. In this scenario, the hot
region detected by Akamatsu \& Kawahara (\citeyear{aka13}) in the
cluster centre could be due to an enhancement of the temperature
related to a three-body merger.

Some hints on the direction of the merger could also be given by the
three head-tail radio galaxies observed in the cluster region (see
Sect.~\ref{notable} and Fig.~\ref{figimage}). Radio sources A and B
lie at the north and point towards the C optical subclump. Considering
that their radial velocities are not extreme and we can see their
radio tails, it is clear that they are moving on the plane of the sky
or, at most, at an intermediate angle. If they are falling into the
cluster following the global direction of the merger, this could be
another clue in support of the feeling that the merging process is
occurring along a direction quite far from the line-of-sight. Similar
considerations could be done also for radio source G, which lies at
the south and is moving inward (but not pointing exactly towards the
cluster centre).

In the context of the scenario described above, we can re-examine the
three hypotheses discussed by Giovannini et al. (\citeyear{gio10}) to
explain the quite complex radio emission of this cluster.  Their major
concern about the hypothesis of two clusters hosting two radio haloes
is that in the southern radio-emitting region there was yet no
evidence of a galaxy overdensity. Our detection of an important
southern subcluster allows now to overcome this concern. However,
Fig.~\ref{figk2z} (lower panel) shows that the north and south radio
emissions are shifted with respect to the centre of the northern and
southern subclusters, at the contrary of the well known case of Abell
399/401 where two radio haloes are hosted in the two pre-merger
clusters (Murgia et al. \citeyear{mur10}). Even if we cannot exclude
that the shifting in ZwCl2341+00 be due to a more advanced but still
pre-merging phase, our above comparison between optical and X-ray data
rather suggests we are in a quite different temporal phase. Moreover,
the detection of polarized flux (quite rare in radio haloes;
Giovannini et al. \citeyear{gio10}) in the north and south
radio-emitting regions cannot be easily reconciled with this scenario.

As for the hypothesis that the radio emission is due to two peripheral
relics and a central radio halo, the presence of the two external
radio emissions at the border of the X-ray emission reminds the
typical clusters with double relics like, e.g., Abell 3667, where two
galaxy subclusters are detected (Roettiger et
al. \citeyear{roe99}). In the case of ZwCl2341+00 the presence of both
relics and halo seems well supported by the existence of, at least,
three subclusters. The central one could be connected to the radio
halo, while the northern and southern subclusters could be responsible
for the relics within a three-body merger scenario. However, we point
out that, if ZwCl2341+00 hosts a central radio halo, its global ICM
temperature ($kT_{\rm X}\sim 4-5$ keV; e.g., Bagchi et
al. \citeyear{bag11}) is an evident anomaly among the X-ray
temperatures (in the range $\sim$7--10 keV) of the clusters hosting
radio haloes included in the DARC sample. At the same time, the
detection of polarization (Giovannini et al. \citeyear{gio10}) also in
the central radio emission casts doubt on the real nature of this
source.

Finally, both the 3D-DEDICA analysis of the spectroscopic sample and
the 2D analysis of the likely members distribution ($r^{\prime}\le
21$) suggest that the cluster and the large-scale structure around it
have an even higher degree of complexity. This can agree with the
hypothesis that the whole region is experiencing a large-scale
structure formation with many shocks originated in a complex multiple
merger. In particular, fig.~3 by Giovannini et al. (\citeyear{gio10})
shows the presence of polarized emission along the east side and they
suggest that it could be related to the direction and dynamics of the
merging process. Indeed, Fig.~\ref{figk2m21} shows several galaxy
groups just in the east region, in particular the very significant NE
group (possibly associated with the X-ray structure Wav-NE),
supporting the hypothesis proposed by Giovannini et
al. (\citeyear{gio10}).

In conclusion, ZwCl2341+00 is a massive galaxy system characterized by
a complex merging process involving several subsystems. Thus, the
detection of diffuse radio emission associated with this cluster is
not surprising evidence. However, the measurement of a larger number
of galaxy redshifts on a wider sky region would be desirable to better
characterize the dynamical properties of the detected subsystems and
to extend the investigation to the large-scale galaxy structure around
the cluster. At the same time, new long exposure time ($\sim$100 ks)
X-ray data would greatly increase our knowledge of the thermal
properties of this cluster and its substructures.

\section*{Acknowledgements}

We are in debt with Gabriele Giovannini for the VLA radio image he
kindly provided us. We also thank the anonymous referee for his/her
useful comments and suggestions. MG acknowledges financial support
from PRIN-INAF2010 and PRIN-MIUR 2010-11 (J91J12000450001). RB
acknowledges financial support from the Spanish Ministry of Economy
and Competitiveness (MINECO) under the 2011 Severo Ochoa Program
MINECO SEV-2011-0187 and the Programa Nacional de Astronom\'ia y
Astrof\'isica AYA2010-21887-C04-04.

This publication is based on observations made on the island of La
Palma with the Italian Telescopio Nazionale Galileo (TNG), which is
operated by the Fundaci\'on Galileo Galilei -- INAF (Istituto
Nazionale di Astrofisica) and is located in the Spanish Observatorio
of the Roque de Los Muchachos of the Instituto de Astrof\'isica de
Canarias.

This research has made use of the galaxy catalogue of the Sloan Digital
Sky Survey (SDSS). The SDSS web site is http://www.sdss.org/, where
the list of the funding organizations and collaborating institutions
can be found.

\input{catalogZwCl2341n1.tex}

\clearpage

\input{tabdedica2d.tex}

\input{tabsub.tex}

\input{tabwavelet.tex}

\end{document}

%% file: catalogZwCl2341n1.tex
%\documentclass[useAMS,usenatbib]{mn2e}
%\usepackage{graphicx}
%\newcommand {\ks} {km~s$^{-1} \;$}
%\newcommand {\kss} {km~s$^{-1}$}
%\newcommand {\mpc} {$Mpc \;$}
%\newcommand {\msun} {$h^{-1} \  M_{\odot} \;$}
%\newcommand {\m} {$M_{\odot} \;$}
%\newcommand {\ml} {$h \, M_{\odot}/L_{\odot} \;$}
%\newcommand {\mll} {$h \, M_{\odot}/L_{\odot}$}
%\newcommand{\vel}{\,{\rm km\,s^{-1}}}
%\newcommand{\tng}{\mathrm{T}}
%\newcommand{\sds}{\mathrm{S}}
%\newcommand{\tns}{\mathrm{T+S}}
%%
%\begin{document}

%\addtocounter{table}{-2}
\begin{table}
        \caption[]{Radial velocities of 128 galaxies in the field of 
	ZwCl2341+00. IDs in italics refer to non-member galaxies.}
         \label{catalogZwCl2341}
              $$ 
        % \begin{array}{p{0.5\linewidth}l}
           \begin{array}{r c r r r}
%            \hline
%            \noalign{\smallskip}
            \hline
            \noalign{\smallskip}

\mathrm{ID} & \alpha,\delta\,(\mathrm{J}2000) & r^{\prime}\,\,\, & v\,\,\,\,\,& \Delta v\\
 & \mathrm{(^h:^m:^s,\degree:':'')}& &\mathrm{(\,km\,s^{-1})}&\mathrm{(\,km\,s^{-1})}\\
            \hline
            \noalign{\smallskip}  

 1 &  23\ 43\ 21.19 ,+00\ 18\ 47.2 &      20.34 &  80403 &  110 \\ 
 2 &  23\ 43\ 21.69 ,+00\ 19\ 33.1 &      17.54 &  80295 &   51 \\ 
 3 &  23\ 43\ 22.48 ,+00\ 18\ 51.2 &      20.21 &  80714 &  114 \\ 
 4 &  23\ 43\ 26.85 ,+00\ 20\ 12.0 &      19.73 &  79868 &  163 \\ 
 5 &  23\ 43\ 26.93 ,+00\ 19\ 49.3 &      19.99 &  80354 &  112 \\ 
 6 &  23\ 43\ 27.13 ,+00\ 18\ 58.8 &      19.85 &  80458 &  119 \\ 
 \textit{7} &  23\ 43\ 28.03 ,+00\ 21\ 29.3 &      18.69 &  20657 &  112 \\ 
 8 &  23\ 43\ 28.18 ,+00\ 19\ 52.5 &      19.90 &  80658 &  145 \\ 
 9 &  23\ 43\ 28.75 ,+00\ 18\ 35.5 &      19.98 &  80543 &  125 \\ 
\textit{10} &  23\ 43\ 29.01 ,+00\ 19\ 49.4 &      18.48 &  39759 &   66 \\ 
11 &  23\ 43\ 30.80 ,+00\ 20\ 44.9 &      20.77 &  82687 &  147 \\ 
12 &  23\ 43\ 31.03 ,+00\ 21\ 36.1 &      18.95 &  81580 &  112 \\ 
\textit{13} &  23\ 43\ 31.50 ,+00\ 20\ 41.8 &      21.80 & 139092 &  119 \\ 
14 &  23\ 43\ 31.61 ,+00\ 19\ 13.4 &      19.63 &  81550 &   79 \\ 
\textit{15} &  23\ 43\ 32.22 ,+00\ 21\ 08.5 &      20.01 &  56247 &  112 \\ 
16 &  23\ 43\ 32.63 ,+00\ 21\ 38.5 &      18.78 &  82414 &   84 \\ 
17 &  23\ 43\ 32.67 ,+00\ 15\ 42.9 &      19.93 &  77584 &   73 \\ 
\textit{18} &  23\ 43\ 32.97 ,+00\ 14\ 57.6 &      20.55 &  55819 &  121 \\ 
19 &  23\ 43\ 33.07 ,+00\ 21\ 27.0 &      18.43 &  80062 &   70 \\ 
20 &  23\ 43\ 34.67 ,+00\ 20\ 37.5 &      17.98 &  80563 &   68 \\ 
21 &  23\ 43\ 35.16 ,+00\ 21\ 27.0 &      19.39 &  82407 &  113 \\ 
22 &  23\ 43\ 35.52 ,+00\ 15\ 02.6 &      19.24 &  81830 &   79 \\ 
23 &  23\ 43\ 35.67 ,+00\ 19\ 51.4 &      18.03 &  80741 &   64 \\ 
\textit{24} &  23\ 43\ 35.69 ,+00\ 20\ 59.9 &      19.74 &  60519 &  117 \\ 
25 &  23\ 43\ 35.81 ,+00\ 21\ 44.8 &      19.54 &  82111 &   99 \\ 
\textit{26} &  23\ 43\ 35.83 ,+00\ 14\ 23.0 &      20.06 &  39810 &  100 \\ 
27 &  23\ 43\ 35.86 ,+00\ 19\ 34.7 &      18.61 &  80893 &   57 \\ 
\textit{28} &  23\ 43\ 36.52 ,+00\ 18\ 39.6 &      19.18 &  56345 &  100 \\ 
\textit{29} &  23\ 43\ 36.97 ,+00\ 21\ 41.2 &      18.78 &  54846 &   84 \\ 
30 &  23\ 43\ 37.48 ,+00\ 17\ 12.7 &      18.38 &  78546 &   48 \\ 
31 &  23\ 43\ 37.84 ,+00\ 17\ 09.7 &      19.26 &  80190 &  103 \\ 
32 &  23\ 43\ 38.26 ,+00\ 19\ 45.6 &      18.64 &  81672 &   73 \\ 
33 &  23\ 43\ 38.27 ,+00\ 18\ 10.8 &      19.64 &  81094 &  108 \\ 
\textit{34} &  23\ 43\ 38.45 ,+00\ 12\ 54.2 &      20.48 &  92302 &  100 \\ 
35 &  23\ 43\ 39.38 ,+00\ 14\ 04.3 &      19.89 &  81166 &   73 \\ 
36 &  23\ 43\ 39.83 ,+00\ 18\ 39.5 &      18.57 &  80964 &   77 \\ 
37 &  23\ 43\ 40.07 ,+00\ 18\ 22.3 &      18.69 &  82195 &  114 \\ 
38 &  23\ 43\ 40.34 ,+00\ 16\ 12.3 &      21.02 &  79235 &   95 \\ 
39 &  23\ 43\ 40.37 ,+00\ 20\ 53.7 &      18.43 &  81415 &   49 \\ 
40 &  23\ 43\ 40.61 ,+00\ 19\ 53.6 &      19.80 &  82692 &  172 \\ 
41 &  23\ 43\ 40.75 ,+00\ 20\ 30.1 &      17.79 &  78291 &   75 \\ 
42 &  23\ 43\ 40.82 ,+00\ 18\ 44.8 &      18.35 &  81480 &   95 \\ 
43 &  23\ 43\ 41.08 ,+00\ 13\ 46.7 &      19.02 &  81456 &   84 \\ 

%            \noalign{\smallskip}			 
%            \hline					    
            \noalign{\smallskip}			    
            \hline					    
         \end{array}					 
     $$ 						 
         \end{table}					 
\addtocounter{table}{-1}				 
\begin{table}					 
          \caption[ ]{Continued.}
     $$ 
           \begin{array}{r c r r r}
%            \hline
%            \noalign{\smallskip}
            \hline
            \noalign{\smallskip}

\mathrm{ID} & \alpha,\delta\,(\mathrm{J}2000) & r^{\prime}\,\,\, & v\,\,\,\,\,& \Delta v\\
 & \mathrm{(^h:^m:^s,\degree:':'')}& &\mathrm{(\,km\,s^{-1})}&\mathrm{(\,km\,s^{-1})}\\

            \hline
            \noalign{\smallskip}

44   &23\ 43\ 41.23 ,+00\ 16\ 35.1 &     20.37&  79179&  103   \\
45   &23\ 43\ 41.24 ,+00\ 16\ 52.7 &     18.34&  80161&   45   \\
46   &23\ 43\ 41.28 ,+00\ 17\ 05.3 &     21.40&  80965&   95   \\
47   &23\ 43\ 41.43 ,+00\ 18\ 21.2 &     19.90&  80286&   52   \\
48   &23\ 43\ 41.57 ,+00\ 19\ 18.6 &     19.19&  79516&  106   \\
49   &23\ 43\ 41.60 ,+00\ 17\ 15.2 &     20.72&  81951&   79   \\
50   &23\ 43\ 41.80 ,+00\ 17\ 26.1 &     20.48&  80620&  158   \\
\textit{51}   &23\ 43\ 41.87 ,+00\ 18\ 03.3 &     19.68&  90420&  132   \\
52   &23\ 43\ 41.94 ,+00\ 17\ 57.3 &     19.69&  81708&   64   \\
53   &23\ 43\ 42.01 ,+00\ 18\ 09.9 &     20.84&  80593&   73   \\
54   &23\ 43\ 42.15 ,+00\ 16\ 14.9 &     20.41&  80440&   77   \\
55   &23\ 43\ 42.49 ,+00\ 19\ 43.9 &     20.53&  80566&  242   \\
56   &23\ 43\ 42.75 ,+00\ 14\ 08.1 &     20.20&  80574&   75   \\
57   &23\ 43\ 43.09 ,+00\ 20\ 27.3 &     19.65&  81954&   97   \\
58   &23\ 43\ 43.55 ,+00\ 11\ 48.0 &     19.70&  80550&   77   \\
59   &23\ 43\ 43.88 ,+00\ 20\ 37.8 &     19.09&  82442&   99   \\
60   &23\ 43\ 43.97 ,+00\ 18\ 04.2 &     18.71&  81621&   73   \\
61   &23\ 43\ 44.03 ,+00\ 20\ 42.7 &     19.30&  82883&  106   \\
62   &23\ 43\ 44.13 ,+00\ 15\ 48.8 &     19.47&  80764&   53   \\
63   &23\ 43\ 44.44 ,+00\ 13\ 01.0 &     20.34&  81970&  123   \\
64   &23\ 43\ 44.61 ,+00\ 15\ 56.6 &     19.65&  77640&  161   \\
65   &23\ 43\ 45.14 ,+00\ 19\ 34.4 &     18.47&  82267&   62   \\
\textit{66}   &23\ 43\ 45.27 ,+00\ 12\ 19.6 &     19.26&  55872&   92   \\
67   &23\ 43\ 45.38 ,+00\ 17\ 40.0 &     19.67&  79886&  121   \\
68   &23\ 43\ 45.39 ,+00\ 14\ 42.2 &     18.95&  81595&   68   \\
\textit{69}   &23\ 43\ 45.50 ,+00\ 14\ 23.4 &     18.47&  55787&   27   \\
70   &23\ 43\ 45.61 ,+00\ 14\ 44.6 &     18.98&  80710&   64   \\
71   &23\ 43\ 45.67 ,+00\ 12\ 45.9 &     20.29&  79842&   57   \\
72   &23\ 43\ 45.97 ,+00\ 15\ 34.3 &     17.93&  77776&   66   \\
73   &23\ 43\ 45.97 ,+00\ 16\ 15.6 &     18.27&  80911&   46   \\
74   &23\ 43\ 46.04 ,+00\ 16\ 46.0 &     19.72&  83112&  101   \\
75   &23\ 43\ 46.17 ,+00\ 15\ 00.4 &     21.23&  79718&  134   \\
76   &23\ 43\ 46.47 ,+00\ 14\ 37.1 &     18.72&  78848&   62   \\
77   &23\ 43\ 46.49 ,+00\ 17\ 24.9 &     19.32&  81893&  128   \\
78   &23\ 43\ 46.61 ,+00\ 17\ 59.0 &     20.67&  81873&  103   \\
\textit{79}   &23\ 43\ 46.85 ,+00\ 11\ 54.8 &     19.73&  92129&   68   \\
80   &23\ 43\ 46.88 ,+00\ 14\ 54.0 &     19.31&  79683&   64   \\
81   &23\ 43\ 47.00 ,+00\ 18\ 07.3 &     19.55&  82501&  110   \\
\textit{82}   &23\ 43\ 47.03 ,+00\ 12\ 47.1 &     18.59&  55884&   40   \\
83   &23\ 43\ 47.10 ,+00\ 16\ 00.2 &     20.01&  79145&   55   \\
84   &23\ 43\ 47.20 ,+00\ 18\ 05.3 &     19.76&  82063&   48   \\
85   &23\ 43\ 47.28 ,+00\ 15\ 20.4 &     19.33&  80988&   66   \\
86   &23\ 43\ 47.30 ,+00\ 10\ 53.2 &     18.92&  81581&   53   \\

%            \noalign{\smallskip}			    
%            \hline					    
            \noalign{\smallskip}			    
            \hline					    
         \end{array}
     $$ 
         \end{table}
\addtocounter{table}{-1}
\begin{table}
          \caption[ ]{Continued.}
     $$ 
           \begin{array}{r c r r r}
%            \hline
%            \noalign{\smallskip}
            \hline
            \noalign{\smallskip}

\mathrm{ID} &\alpha,\delta\,(\mathrm{J}2000) & r^{\prime} & v\,\,\,\,\,& \Delta v\\
 & \mathrm{(^h:^m:^s,\degree:':'')}& &\mathrm{(\,km\,s^{-1})}&\mathrm{(\,km\,s^{-1})}\\

            \hline
            \noalign{\smallskip}
   
 87   &234347.31 ,+001656.0 &     18.19&  78104&   51  \\
 88   &234347.37 ,+001607.7 &     20.04&  79728&  169  \\
 89   &234347.42 ,+001953.6 &     18.28&  80505&   92  \\
\textit{90}   &234347.48 ,+001152.1 &     20.27&  92295&   86  \\
 91   &234347.53 ,+001524.1 &     17.94&  79987&   57  \\
\textit{92}   &234347.57 ,+001421.8 &     20.79& 115648&  108  \\
 93   &234347.66 ,+001548.7 &     20.85&  80907&  117  \\
 94   &234347.81 ,+001633.9 &     19.13&  77771&   51  \\
 95   &234347.96 ,+001401.2 &     19.75&  78098&   86  \\
 96   &234347.97 ,+001556.2 &     19.49&  78852&   99  \\
 97   &234348.14 ,+001846.1 &     20.23&  81885&   90  \\
 98   &234348.15 ,+001414.5 &     20.72&  82260&   88  \\
 99   &234348.29 ,+002010.2 &     19.71&  80662&  108  \\
100   &234348.49 ,+001248.7 &     19.09&  79462&   51  \\
101   &234348.53 ,+001507.6 &     18.37&  81211&   59  \\
102   &234349.04 ,+001548.5 &     19.39&  80407&   51  \\
103   &234349.08 ,+001537.6 &     19.51&  77722&   44  \\
104   &234349.29 ,+002017.5 &     20.00&  80021&   90  \\
105   &234349.40 ,+001606.7 &     21.17&  78603&  114  \\
106   &234350.74 ,+001547.4 &     -.-  &  79726&  100  \\
107   &234350.85 ,+002005.3 &     19.70&  80819&   92  \\
108   &234351.15 ,+001137.0 &     19.31&  81276&   59  \\
\textit{109}  &234351.45 ,+001718.1 &     21.12& 156705&   95  \\
\textit{110}  &234352.25 ,+001121.5 &     20.10& 136529&   57  \\
111   &234352.25 ,+002045.8 &     20.47&  81693&  145  \\
\textit{112}  &234352.47 ,+002018.9 &     19.79& 116221&  176  \\
\textit{113}  &234352.73 ,+001708.8 &     21.10& 191834&  112  \\
114   &234353.25 ,+001725.9 &     20.07&  80442&  141  \\
115   &234353.84 ,+001716.7 &     20.43&  80126&   99  \\
116   &234354.59 ,+001728.1 &     18.27&  81141&   73  \\
117   &234354.61 ,+001450.8 &     19.99&  78984&   59  \\
\textit{118}  &234354.87 ,+001514.6 &     22.17&  69995&  164  \\
119   &234355.01 ,+001436.3 &     20.03&  81931&   53  \\
120   &234355.79 ,+001425.9 &     19.03&  82372&   40  \\
\textit{121}  &234355.82 ,+001723.8 &     18.47&  41282&   92  \\
\textit{122}  &234355.97 ,+001229.2 &     20.17&  92448&   97  \\
123   &234357.12 ,+001723.2 &     19.36&  79778&   81  \\
\textit{124}  &234358.57 ,+001459.2 &     18.13&  47167&  100  \\
\textit{125}  &234400.77 ,+001720.6 &     20.58&  55476&  106  \\
126   &234401.02 ,+001555.3 &     19.80&  80331&   99  \\
\textit{127}  &234401.54 ,+001723.8 &     21.34& 159682&  100  \\
128   &234404.61 ,+001456.8 &     18.35&  81291&   53  \\
                                                     
%            \noalign{\smallskip}			    
%            \hline					    
            \noalign{\smallskip}			    
            \hline					    
         \end{array}
     $$ 
\end{table}

%\end{document}

%% file: tabdedica2d.tex
\begin{table*}
        \caption[]{2D substructure from the SDSS photometric samples.}
         \label{tabdedica2d}
            $$
         \begin{array}{l r c c r }
%            \hline
%            \noalign{\smallskip}
            \hline
            \noalign{\smallskip}
\mathrm{Subclump} & N_{\rm S} & \alpha({\rm J}2000),\,\delta({\rm J}2000)&\rho_{
\rm S}&\chi^2_{\rm S}\\
& & \mathrm{(^h:^m:^s,\degree:':'')}&&\\
         \hline
         \noalign{\smallskip}
\mathrm{C\ (}r^{\prime}\le 20)           & 20&23\ 43\ 41.7,+00\ 18\ 13&1.00&16\\
\mathrm{SSE\ (}r^{\prime}\le 20)         & 31&23\ 43\ 47.8,+00\ 15\ 20&0.77&15\\
\mathrm{NNW\ (}r^{\prime}\le 20)         & 23&23\ 43\ 36.9,+00\ 19\ 52&0.55&12\\
\mathrm{N-NNW\ (}r^{\prime}\le 20)       & 10&23\ 43\ 33.3,+00\ 21\ 29&0.50& 8\\
\noalign{\smallskip}
%\hline
\mathrm{C1\ (}r^{\prime}\le 21)          & 12&23\ 43\ 42.4,+00\ 18\ 09&1.00&10\\
\mathrm{C2\ (}r^{\prime}\le 21)          & 10&23\ 43\ 45.9,+00\ 17\ 46&0.89& 7\\
\mathrm{NNW\ (}r^{\prime}\le 21)         & 18&23\ 43\ 36.3,+00\ 19\ 47&0.79&10\\
\mathrm{SSE\ (}r^{\prime}\le 21)         & 15&23\ 43\ 47.7,+00\ 15\ 51&0.77&10\\
\mathrm{NE\ (}r^{\prime}\le 21)          & 14&23\ 43\ 51.7,+00\ 20\ 25&0.63&10\\
\mathrm{ext\ SSW^{\mathrm{a}}\ (}r^{\prime}\le 21)&  7&23\ 42\ 58.5,-00\ 10\ 35&0.56&11\\
\mathrm{ext\ SSE^{\mathrm{b}}\ (}r^{\prime}\le 21)& 17&23\ 44\ 46.3,-00\ 05\ 11&0.51& 9\\
\mathrm{N-NNW\ (}r^{\prime}\le 21)       & 14&23\ 43\ 31.7,+00\ 21\ 35&0.50& 7\\

%              \noalign{\smallskip}
              \noalign{\smallskip}
%            \hline
%            \noalign{\smallskip}
            \hline
         \end{array}
$$
\begin{list}{}{}
\item[$^{\mathrm{a}}$] Corresponding to the galaxy cluster SDSS CE J355.744965--00.175613 ($z \sim 0.25$; see Goto et al. \citeyear{got02}).
\item[$^{\mathrm{b}}$] Corresponding to the galaxy cluster SDSS CE J356.199097--00.084485 ($z \sim 0.23$; see Goto et al. \citeyear{got02}).
\end{list}

\end{table*}

%% file: tabsub.tex
\begin{table*}
        \caption[]{Kinematical properties of the whole cluster and individual subclusters.}
         \label{tabsub}
            $$
         \begin{array}{l r r r l l}
            \hline
            \noalign{\smallskip}
\mathrm{System} & N_{\rm g} &<v>&\sigma_{V}&R_{\rm vir}&\mathrm{M(}<R_{\rm vir})\\
& &\mathrm{(km\ s^{-1})}&\mathrm{(km\ s^{-1})}&\mathrm{(Mpc)}&(10^{14}\mathrm{M}_{\sun})\\
         \hline
         \noalign{\smallskip}
\mathrm{Whole\ system}   &101& 80726\pm103& 1034_{\ 74}^{\ 88}& 2.2&15_{-4}^{+5}\\
         \noalign{\smallskip}
%         \hline
\mathrm{KMM1D-LV}        &  7& 77827^{\mathrm{a}}      & 152^{\mathrm{a}} &\ 0.32& 0.05      \\
\mathrm{KMM1D-HV}        & 94& 80843^{\mathrm{a}}      & 857^{\mathrm{a}}&\ 1.8&  8.7        \\
          \noalign{\smallskip}
%          \hline
\mathrm{KMM3D-S-LV}      & 49& 80046\pm154& 1069_{\ 84}^{106}&2.3& 17\pm5 \\ 
\mathrm{KMM3D-C-HV}      & 22& 81213\pm186&  848_{\ 97}^{216}&1.8&\ 8_{-3}^{+5} \\ 
\mathrm{KMM3D-N-HV}      & 30& 81316\pm136&  733_{\ 54}^{\ 69}&1.6&\ 5\pm2\\ 
        \noalign{\smallskip} 
%         \hline
\mathrm{HT-S-LV}         &  6& 78172\pm517& 1066_{377}^{445}&2.3& 17_{-13}^{+15} \\
\mathrm{HT-S-MV}         & 14& 80293\pm228&  810_{179}^{267}&1.7&\ 7_{-4}^{+5}\\
\mathrm{HT-NC-HV}        & 23& 81139\pm138&  642_{\ 62}^{108}& 1.4&\ 4_{-1}^{+2}\\
         \noalign{\smallskip}
%         \hline
\mathrm{DED3D-S-LLV}     &  7& 77795\pm67 &  156_{\ 87}^{\ 46}&0.33&\ 0.05_{-0.06}^{+0.04} \\
\mathrm{DED3D-S-LV}      & 21& 79486\pm126&  559_{\ 73}^{101}&1.2&\ 2_{-0.9}^{+1} \\
\mathrm{DED3D-NCS-MV}    & 49& 80723\pm69 &  479_{\ 62}^{\ 73}&1.0&\ 1.5_{-0.5}^{+0.6} \\
\mathrm{DED3D-NC-HV}     & 19& 82006\pm89 &  372_{\ 60}^{\ 85}&0.79&\ 0.7_{-0.3}^{+0.4} \\
\mathrm{DED3D-N-HHV}     &  5& 82273\pm199&  339_{324}^{134}&0.72&\ 0.5_{-1.0}^{+0.4}  \\
              \noalign{\smallskip}
%             \noalign{\smallskip}
            \hline
%            \noalign{\smallskip}
%            \hline
         \end{array}
$$
\begin{list}{}{}  
\item[$^{\mathrm{a}}$] These quantities are computed weighting galaxies
  according to their partial membership to both the 1D-KMM groups (see
  text). For other subclusters we use the usual robust estimates on
  group members.
\end{list}

\end{table*}

%% file: tabwavelet.tex
\begin{table*}
        \caption[]{Substructure from the wavelet analysis of {\it Chandra} data.}
         \label{tabwavelet}
            $$
         \begin{array}{l c c}
%            \hline
%            \noalign{\smallskip}
            \hline
            \noalign{\smallskip}
\mathrm{Subclump} & \alpha({\rm J}2000),\,\delta({\rm J}2000)&\mathrm{Significance^a}\\
 & \mathrm{(^h:^m:^s,\degree:':'')}& \\
         \hline
         \noalign{\smallskip}
\mathrm{Wav-NNW} &23\ 43\ 37.2,+00\ 19\ 33&14.74\\
\mathrm{Wav-C}   &23\ 43\ 41.0,+00\ 18\ 03& 8.72\\
\mathrm{Wav-SSE} &23\ 43\ 44.8,+00\ 15\ 26& 6.01\\
\mathrm{Wav-NE}  &23\ 43\ 46.8,+00\ 20\ 14& 5.73\\
              \noalign{\smallskip}
%              \noalign{\smallskip}
            \hline
%            \noalign{\smallskip}
%            \hline
         \end{array}
$$
\begin{list}{}{}
\item[$^{\mathrm{a}}$] Significance estimate in units of $\sigma$,
  computed by dividing the net source counts by the ``Gehrels error''
  $\sigma_{\rm G}$ of the background counts in the source region (see
  {\sevensize CIAO} manual for details).
\end{list}

\end{table*}

%% file: zwcl2341_astroph.bbl
\begin{thebibliography}{99}

\bibitem[\protect\citeauthoryear{Akamatsu \& Kawahara}{2013}]{aka13} Akamatsu H., Kawahara H., 2013, PASJ, 65, 16

\bibitem[\protect\citeauthoryear{Ashman et al.}{1994}]{ash94} Ashman K. M., Bird C. M., Zepf S. E., 1994, AJ, 108, 2348

\bibitem[\protect\citeauthoryear{Bagchi et al.}{2002}]{bag02} Bagchi J., Ensslin T. A., Miniati F., Stalin C. S., Singh M., Raychaudhury S., Humeshkar N. B., 2002, New Astron., 7, 249

\bibitem[\protect\citeauthoryear{Bagchi et al.}{2011}]{bag11} Bagchi J., et al., 2011, Mem. Soc. Astron. Ital., 82, 561

\bibitem[\protect\citeauthoryear{Barrena et al.}{2011}]{bar11} Barrena R., Girardi M., Boschin W., de Grandi S., Eckert D., Rossetti M., 2011, A\&A, 529, A128

\bibitem[\protect\citeauthoryear{Beers et al.}{1990}]{bee90} Beers T. C., Flynn K., Gebhardt K., 1990, AJ, 100, 32

\bibitem[\protect\citeauthoryear{Beers et al.}{1991}]{bee91} Beers T. C., Forman W., Huchra J. P., Jones C., Gebhardt K., 1991, AJ, 102, 1581

\bibitem[\protect\citeauthoryear{Beers et al.}{1992}]{bee92} Beers T. C., Gebhardt K., Huchra J. P., Forman W., Jones C., Bothun G. D., 1992, ApJ, 400, 410

\bibitem[\protect\citeauthoryear{Bird}{1994}]{bir94} Bird C. M., 1994, AJ, 107, 1637

\bibitem[\protect\citeauthoryear{Bird \& Beers}{1993}]{bir93} Bird C. M., Beers, T. C., 1993, AJ, 105, 1596

\bibitem[\protect\citeauthoryear{Boschin et al.}{2004}]{bos04} Boschin W., Girardi M., Barrena R., Biviano A., Feretti L., Ramella M., 2004, A\&A, 416, 839

\bibitem[\protect\citeauthoryear{Boschin et al.}{2012a}]{bos12a} Boschin W., Girardi M., Barrena R., Nonino M., 2012a, A\&A, 540, A43

\bibitem[\protect\citeauthoryear{Boschin et al.}{2012b}]{bos12b} Boschin W., Girardi M., Barrena R., 2012b, A\&A, 547, A44

\bibitem[\protect\citeauthoryear{Brown \& Rudnick}{2009}]{bro09} Brown S., Rudnick L., 2009, AJ, 137, 3158

\bibitem[\protect\citeauthoryear{Brunetti et al.}{2009}]{bru09} Brunetti G., Cassano R., Dolag K., Setti G., 2009, A\&A, 507, 661

\bibitem[\protect\citeauthoryear{Cassano et al.}{2006}]{cas06} Cassano R., Brunetti G., Setti G., 2006, MNRAS, 369, 1577

\bibitem[\protect\citeauthoryear{den Hartog \& Katgert}{1996}]{den96} den Hartog R., Katgert P., 1996, MNRAS, 279, 349

\bibitem[\protect\citeauthoryear{Dressler}{1980}]{dre80} Dressler A., 1980, ApJ, 236, 351 

\bibitem[\protect\citeauthoryear{Dressler \& Schectman}{1988}]{dre88} Dressler A., Shectman S. A., 1988, AJ, 95, 985

\bibitem[\protect\citeauthoryear{Ensslin et al.}{1998}]{ens98} Ensslin T. A., Biermann P. L., Klein U., Kohle S., 1998, A\&A, 332, 395

\bibitem[\protect\citeauthoryear{Fadda et al.}{1996}]{fad96} Fadda D., Girardi M., Giuricin G., Mardirossian F., Mezzetti M., 1996, ApJ, 473, 670

\bibitem[\protect\citeauthoryear{Fasano et al.}{1987}]{fas87} Fasano G., Franceschini A., 1987, MNRAS, 225, 155

\bibitem[\protect\citeauthoryear{Feretti}{1999}]{fer99} Feretti L., 1999, MPE Rep., 271, 3

\bibitem[\protect\citeauthoryear{Feretti et al.}{2012}]{fer12} Feretti L., Giovannini G., Govoni F., Murgia M., 2012, A\&AR, 20, 54

\bibitem[\protect\citeauthoryear{Ferrari et al.}{2003}]{fer03} Ferrari C., Maurogordato S., Cappi A., Benoist C., 2003, A\&A, 399, 813

\bibitem[\protect\citeauthoryear{Ferrari et al.}{2008}]{fer08} Ferrari C., Govoni F., Schindler S., Bykov A. M., Rephaeli Y., 2008, Space Sci. Rev., 134, 93

\bibitem[\protect\citeauthoryear{Giovannini et al.}{2010}]{gio10} Giovannini G., Bonafede A., Feretti L., Govoni F., Murgia M., 2010, A\&A, 511, L5

\bibitem[\protect\citeauthoryear{Girardi \& Mezzetti}{2001}]{gir01} Girardi M., Mezzetti M., 2001, ApJ, 548, 79

\bibitem[\protect\citeauthoryear{Girardi et al.}{1996}]{gir96} Girardi M., Fadda D., Giuricin G., Mardirossian F., Mezzetti M., Biviano A., 1996, ApJ, 457, 61

\bibitem[\protect\citeauthoryear{Girardi et al.}{1997}]{gir97} Girardi M., Escalera E., Fadda D., Giuricin G., Mardirossian F., Mezzetti M., 1997, ApJ, 482, 11

\bibitem[\protect\citeauthoryear{Girardi et al.}{1998}]{gir98} Girardi M., Giuricin G., Mardirossian F., Mezzetti M., Boschin W., 1998, ApJ, 505, 74

\bibitem[\protect\citeauthoryear{Girardi et al.}{2010a}]{gir10conf} Girardi M., Barrena R., Boschin W., 2010a, in `Galaxy Clusters: Observations, Physics and Cosmology'. Published online at the site http://www.mpa-garching.mpg.de/$\sim$clust10/

\bibitem[\protect\citeauthoryear{Girardi et al.}{2010b}]{gir10} Girardi M., Boschin W., Barrena R., 2010b, A\&A, 517, A65

\bibitem[\protect\citeauthoryear{Girardi et al.}{2011}]{gir11} Girardi M., Bardelli S., Barrena R., Boschin W., Gastaldello F., Nonino M., 2011, A\&A, 536, A89

\bibitem[\protect\citeauthoryear{Goto et al.}{2002}]{got02} Goto T. et al., 2002, AJ, 123, 1807

\bibitem[\protect\citeauthoryear{Govoni et al.}{2001}]{gov01} Govoni F., Ensslin T. A., Feretti L., Giovannini G., 2001, A\&A, 369, 441

\bibitem[\protect\citeauthoryear{Hoeft et al.}{2004}]{hoe04} Hoeft M., Br\"uggen M., Yepes G., 2004, MNRAS, 347, 389

\bibitem[\protect\citeauthoryear{Kim et al.}{1989}]{kim89} Kim K.-T., Kronberg P. P., Giovannini G., Venturi T., 1989, Nature, 341, 720

\bibitem[\protect\citeauthoryear{Maurogordato et al.}{2011}]{mau11} Maurogordato S., Sauvageot J. L., Bourdin H., Cappi A., Benoist C., Ferrari C., Mars G., Houairi K., 2011, A\&A, 525, A79

\bibitem[\protect\citeauthoryear{Murgia et al.}{2010}]{mur10} Murgia M., Govoni F., Feretti L., Giovannini G., 2010, A\&A, 509, A86

\bibitem[\protect\citeauthoryear{Pisani}{1993}]{pis93} Pisani A., 1993, MNRAS, 265, 706

\bibitem[\protect\citeauthoryear{Pisani}{1996}]{pis96} Pisani A., 1996, MNRAS, 278, 697

\bibitem[\protect\citeauthoryear{Pizzo et al.}{2008}]{piz08} Pizzo R. F., de Bruyn A. G., Feretti L., Govoni F., 2008, A\&A, 481, L91

\bibitem[\protect\citeauthoryear{Roettiger et al.}{1999}]{roe99} Roettiger K., Burns J. O., Stone J. M., 1999, ApJ, 518, 603

\bibitem[\protect\citeauthoryear{Serna \& Gerbal}{1996}]{ser96} Serna A., Gerbal D., 1996, A\&A, 309, 65

\bibitem[\protect\citeauthoryear{Serra \& Diaferio}{2013}]{ser13} Serra A. L., Diaferio A., 2013, ApJ, 768, 116

\bibitem[\protect\citeauthoryear{Shapiro \& Silk}{1965}]{sha65} Shapiro S. S., Wilk M. B., 1965, Biometrika, 52, 591

\bibitem[\protect\citeauthoryear{Tonry \& Davis}{1979}]{ton79} Tonry J., Davis M., 1979, AJ, 84, 1511

\bibitem[\protect\citeauthoryear{van Weeren et al.}{2009a}]{wer09a} van Weeren R. J. et al., 2009a, A\&A, 506, 1083

\bibitem[\protect\citeauthoryear{van Weeren et al.}{2009b}]{wer09b} van Weeren R. J., Intema H. T., Oonk J. B. R., R\"ottgering H. J. A., Clarke T. E., 2009b, A\&A, 508, 1269

\bibitem[\protect\citeauthoryear{Venturi}{2011}]{ven11} Venturi T., 2011, Mem. Soc. Astron. Ital., 82, 499

\end{thebibliography}
